 \documentclass{jfm}
 \usepackage{setspace}
 \usepackage{graphicx}
 \usepackage{epstopdf, epsfig}
 \usepackage{amsmath,amssymb} 
 \usepackage{subfig}
 \usepackage{amsfonts}
 \usepackage{pgfplots}
 \pgfplotsset{compat=1.5}
 \usepackage{caption}
 \usepackage{floatrow}
 \usepackage{ragged2e}
 \usepackage{soul}
 \usepackage{xcolor}
 
 \def\IB#1{\boldsymbol{#1}} 
 \def\W/!i#1{\Wi} 
 \def\bten#1{\IB{\mathsfbi{#1}}}

 \shorttitle{Electrokinetically enhanced cross-stream particle migration in viscoelastic flows}
 \shortauthor{A. Choudhary, D. Li, T. Renganathan, X. Xuan and S. Pushpavanam}
 
 \title{Electrokinetically enhanced cross-stream particle migration in viscoelastic flows}
 
 \author{Akash Choudhary\aff{1},
 	Di Li\aff{2},
 	T. Renganathan\aff{1},
 	Xiangchun Xuan\aff{2}$\dagger$,
 	\and S. Pushpavanam\aff{1}
 	\corresp{\email{xcxuan@clemson.edu,  spush@iitm.ac.in}}
 }
 
 \affiliation{\aff{1}Department of Chemical Engineering, Indian Institute of Technology Madras,TN 600036, India \\ \aff{2}Department of Mechanical Engineering, Clemson University, South Carolina 29634-0921, USA.}
 
\begin{document}

 	\maketitle
 	
 	\begin{abstract}
 		Advancements in understanding the lateral migration of particles have helped in enhanced focusing in microfluidic devices.
 		In this work, we investigate the effects of electrokinetics on particle migration in a viscoelastic flow, where the electric field is applied parallel to the flow.
 		Through experiments and use of perturbation theory in conjunction with the reciprocal theorem, we show that the interaction of electrokinetic and rheological effects can result in an enhancement in migration by an order of magnitude.
 		The theoretical analysis, in agreement with the experiments, demonstrates that the particles can be focused at different equilibrium positions based on their intrinsic electrical properties.
 	\end{abstract}

 	\section{Introduction}
 	In the past two decades, biological and healthcare research has significantly benefited by the application of various particle and cell sorting techniques---inertial focusing \citep{di2009inertial,warkiani2015membrane,zhang2016} and viscoelastic focusing \citep{leshansky2007tunable,xuan2017review}---in microfluidic devices.
 	The working principle of such devices is based on the exploitation of non-linearity in the flow (arising from inertial or polymeric stresses) which  breaks the symmetry and generates a lateral force on the particle, pushing it towards the equilibrium positions.
 	Since most biological fluids are non-Newtonian, it is essential to understand the rheological effects on particle migration. The pioneering studies of Mason and his coworkers \citep{karnis1966particle,gauthier1971particle} and \cite{ho1976migration} clearly demonstrated that the presence of polymer chains in the fluid exert a normal stress on the particle, pushing it towards the region of lowest shear. In biological fluids, strands of DNA enact the role of polymeric chains and help in the lateral migration of cells \citep{kim2016elasto}.
 	In recent years, this normal stress induced migration has been exploited to design tubular-viscoelastic-focusers, in which the region of lowest shear is the channel axis \citep{avinoAnnRev}. 
 	However, the labor-intensive fabrication process limits the accessibility of these devices \citep{nam2015hybrid}.
 	The more readily available rectangular microchannels are plagued by the limitation of multiple equilibrium positions (1 center and 4 corners), which renders the downstream detection and separation difficult \citep{xuan2017review,tian2019manipulation}. Furthermore, the passive nature of the migration requires large focusing lengths which are dependent on the flowrate \citep{avinoAnnRev}. 
 	
 	Unlike flows at large scale, in micro-scale flows, the short-ranged effects near the particle surface can determine their fate \citep{datt2017activeComplex}. One can resort to manipulation of surface effects through external fields to tune particle focusing \citep{rossi2019particle}. 
 	This has potential applications in focusing, separation and detection of rare circulating tumor cells in the blood stream \citep{zhang2014real}. 
 	Recent studies have shown that cancer cells exhibit highly negative charge on their surface which can be used to separate them from the blood stream by applying external electric fields \citep{cancer_negative_Chen,cancer_2019}. 
 	In our previous study \citep{choudhary2019inertial}, we studied the effects of electric fields on the inertial migration in pressure driven flow of a Newtonian fluid.
 	Since most biological fluids exhibit non-Newtonian behaviour, it is important  to understand the influence of electrokinetics on the migration in complex fluids.
 	

 	In this work, we experimentally and theoretically investigate the influence of electrophoresis on viscoelastic focusing of particles, where the electric field is applied parallel to the flow. 
 	A vast majority of cells and particles, when submerged in aqueous electrolytes, acquire a surface charge which attracts the counter-charged ions. As a result a thin double layer forms around the particle \citep{anderson1989}. 
 	An external electric field causes a tangential slip on the surface of the particle. This generates a disturbance field around the particle which may amplify the asymmetry in normal stresses.
 	Our experimental results demonstrate that the focusing is significantly accelerated in the presence of an electric field, which reduces the overall focusing length. Furthermore, the electric field provides an external control over the particle migration and enables charge-selective focusing.
 	We explain the experimental observations qualitatively and quantitatively using an approach based on perturbation theory:
 	we analyze the system in the limit of slow and slowly varying flows and capture the viscoelastic effects using a second order fluid  model \citep{ho1976migration,bird1987dynamics}. This allows us to predict the migration velocity through a perturbation solution. 
 	Using the reciprocal theorem we show that the electrophoretic motion of the particle generates additional polymeric stresses which enhance the migration by $ O(\kappa^{-1}) $, where $ \kappa $ is the particle to channel size ratio.
 	The theoretical results (obtained without having to resort to fit any parameter) are found to be in reasonable agreement with the experimental observations.
 	
 	The rest of this article is organized as follows. In \S 2 we describe the experimental conditions and results. The explanation of these results is conducted through a mathematical model which is elaborated in \S 3. In \S 4 we qualitatively and quantitatively compare the theoretical and experimental results. Key conclusions and potential future directions are described in \S 5.


 	\section{Experiments}
 	In our experiments, we use a 2 cm long PDMS microchannel with a rectangular cross-section of $66 \times 54$ $ \mu $m$ ^{2} $. 
 	The electric fields were generated using a high-voltage DC power supply using a wire connection to electrodes at the channel inlet and outlet. 
 	We dissolved polyethylene oxide powder (PEO, molecular weight = 2$\times 10^{6}$ g/mol, Sigma-Aldrich)  into a buffer solution (0.01 mM phosphate buffer mixed with 0.5$ \% $ Tween 20, Fisher Scientific) at a concentration of 250 ppm.
 	A dilute suspension ($ < $0.1$ \% $ in volume fraction) of polystyrene spheres of 2.2 $ \mu $m diameter is prepared in the PEO solution and is pumped at 25 $  \rm{\mu L/ h} $. 
 	The particle zeta potential was measured to be $ -83 $ mV. 
 	Particle motion was visualized at the outlet of the microchannel through an inverted microscope (Nikon Eclipse TE2000U) equipped with a CCD camera (Nikon DS-Qi1MC). Digital videos were recorded at a rate of 15 frames per second, from which the superimposed images were obtained and further processed in the Nikon imaging software (NIS-Elements AR 3.22). 

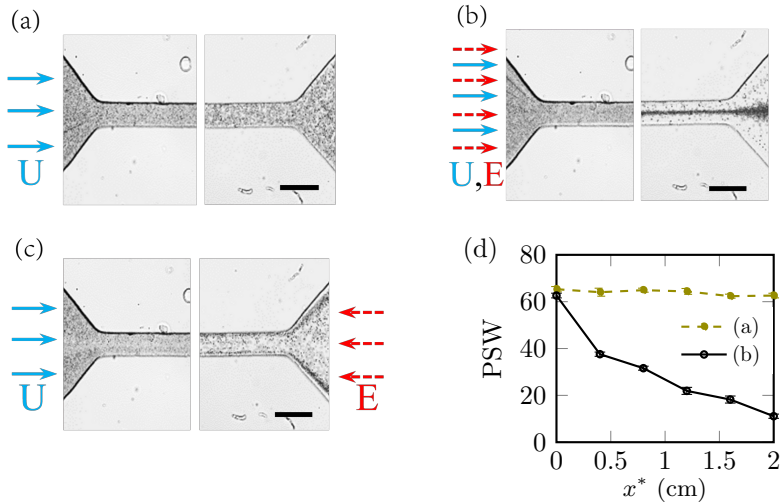
\begin{figure*}%
 		\centering
 		{{\includegraphics[scale=1.14]{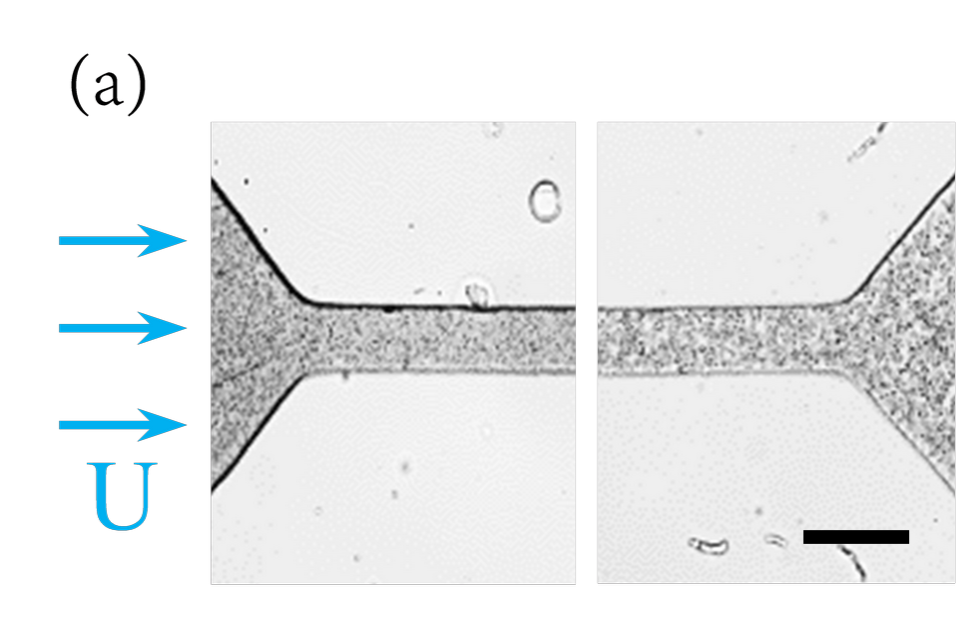} }}%
 		\; \; \; \; \;
 		{{\includegraphics[scale=1.1]{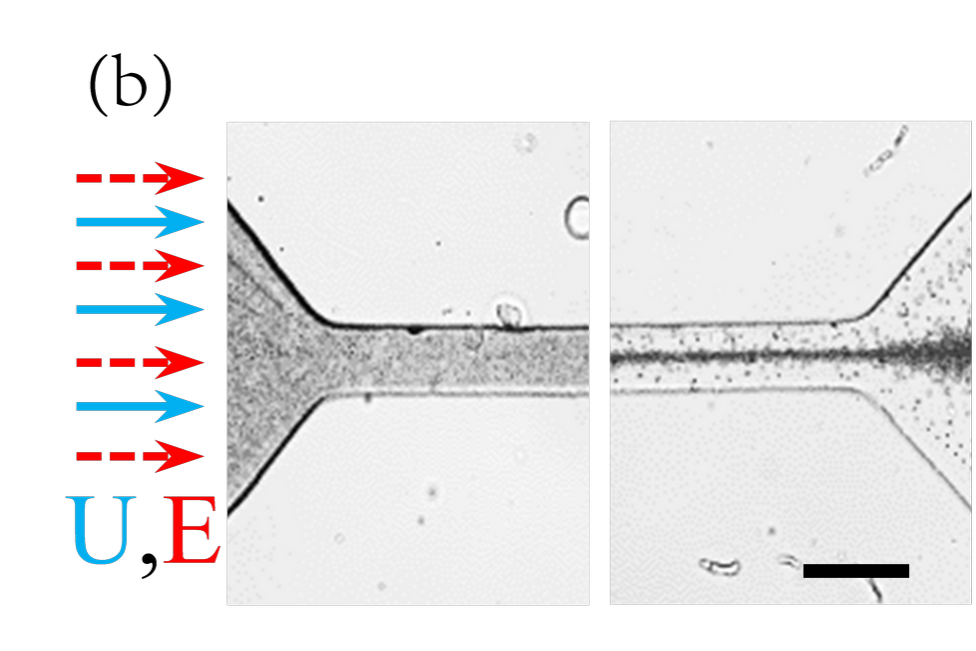} }}%
 		\; \; 
 		{{\includegraphics[scale=1.1]{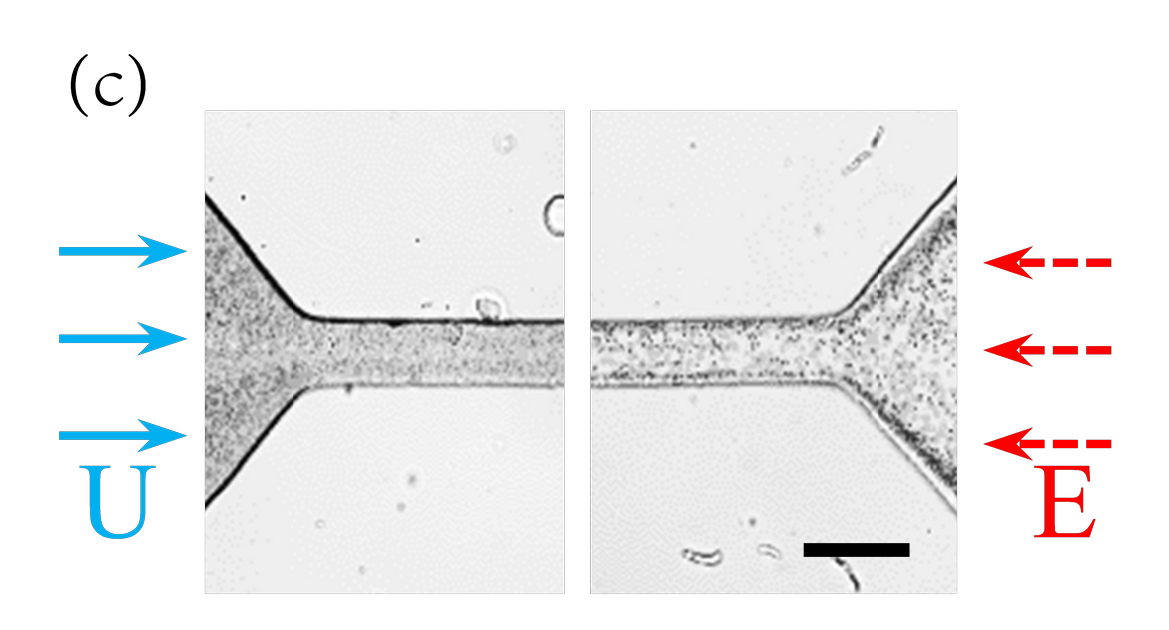} }}
 		\; \;
 		\begin{tikzpicture}[baseline]
 		\begin{axis}[clip=false,
 		width=0.33\textwidth, height=0.30\textwidth,
 		xlabel= {\small $ x^{*} $ (cm)}, ylabel=PSW, 
 		ylabel near ticks,
 		xlabel shift= -4 pt,
 		xmin=0, xmax=2 , ymin=0, ymax=80, thick, legend style={draw=none,at={(1,0.36)},anchor=south east},	legend style={nodes={scale=0.8, transform shape}
 			,
 			legend columns=1, 
 			,	/tikz/column 2/.style={
 				column sep=5pt,
 			},
 		},]
 		
 		\addplot[olive ,dashed, mark=*, mark options={scale=0.6}, error bars/.cd, y dir=both, y explicit] table[x=Len, y=PSW, y error= err] {
 			Len	PSW	err
 			0	65.33	1.1282287
 			0.4	64.05	1.705002444
 			0.8	64.91	0.855004873
 			1.2	64.39	1.239771484
 			1.6	62.35	0.855004873
 			2	62.61	0.989764282
 			
 		};
 		
 		\addplot[black, mark=o, mark options={scale=0.6}, error bars/.cd, y dir=both, y explicit] table[x=Len, y=PSW, y error= err] {
 			Len	PSW	err
 			0	62.62	1
 			0.4	37.58	0.9
 			0.8	31.6	0.85
 			1.2	21.92	1.5
 			1.6	18.22	1.5
 			2	11.1	0.9
 		};
 		
 		\legend{(a), (b)}
 		\node[scale=1] at (axis cs: -0.7, 82) {(d)}; 
 		\end{axis}
 		\end{tikzpicture}
 		\caption{Superimposed experimental images showing the focusing of polystyrene particles (negative zeta potential), at the channel inlet and outlet, for: (a) no external electric field, (b) electric field in the direction of the flow. (c) Particle focusing when electric field is applied opposite to the flow. (d) Particle stream width for (a) and (b). Scale bar represents $ 100 \, \mu $m.
 			Parameters:  $ \mathcal{E}_{\infty}^{*}=300 $ V/cm, $ C^{*}_{\rm{PEO}} =250 $ ppm, $ a^{*}=1.1 $ $ \mu $m, $ l^{*}= 66 \, \rm{\mu} $m, $ U_{max}^{*}=2.9 $ mm/s, $ \mu^{*}=1.3  $ mPa.s, $ \zeta_{p}^{*}=- 83 $ mV, $ \zeta_{w}^{*}=-120 $ mV, $ Re_{p}=1.6 \times 10^{-4} $. Experimental movies corresponding the above superimposed images can be found in the supplementary material. }
 		\label{fig:exp}%
 	\end{figure*}

 	Fig.\ref{fig:exp}(a) shows the particles at the inlet and outlet of the channel when no electric field is applied \textit{i.e.,} the focusing is purely viscoelastic \citep{karnis1966particle,gauthier1971particle,ho1976migration}. The particles remain almost uniformly distributed at the channel exit.
 	This weak focusing is depicted in Fig.\ref{fig:exp}(d), which shows the change in particle stream width (lateral width of the particle distribution) along the channel length.
 	In Fig.\ref{fig:exp}(b) an application of electric field parallel to the flow shows that a significant fraction of the particles are focused at the centerline. 
 	Fig.\ref{fig:exp}(d) shows that the introduction of electric field results in substantial reduction in particle stream width.
 	Upon reversing the direction of the electric field (see Fig.\ref{fig:exp}(c)), the particles focus near the walls.
 	These results demonstrate the potential to efficiently separate particles and cells based on their intrinsic electrical properties. 
 	To extrapolate these observations towards potential applications, it is pivotal to understand the role of electrical properties and flow profile characteristics.

 	In what follows, we formulate a mathematical framework to predict, quantify and draw physical insights into the migration behavior observed in the experiments. Since the particle suspension in the experimental study is dilute, we analyze the migration of a single particle.
 	We model the fluid medium using a second-order fluid (SOF) model \citep{bird1987dynamics}, as we are only interested in capturing the effect of small deviations from the Newtonian behavior. 
 	Although the SOF model is only valid for slow and slowly varying (thus nearly Newtonian) flows, it's an asymptotic approximation for most viscoelastic flows \citep{ho1976migration}.
 	Furthermore, it has been shown in the past that dilute PEO solutions ($ C_{PEO}^{*} < 500 $ ppm) exhibit pure elastic behavior (i.e. negligible shear thinning) \citep{rodd2005inertio,yang2011sheathless,kim2012lateral}. Hence employing the SOF model can help us capture the dynamics in slow Boger fluid flows, both qualitatively and quantitatively.

 	\floatsetup[figure]{style=plain}

 	\section{Mathematical model}
 	
\begin{figure}
 		\centerline{\includegraphics[scale=0.36]{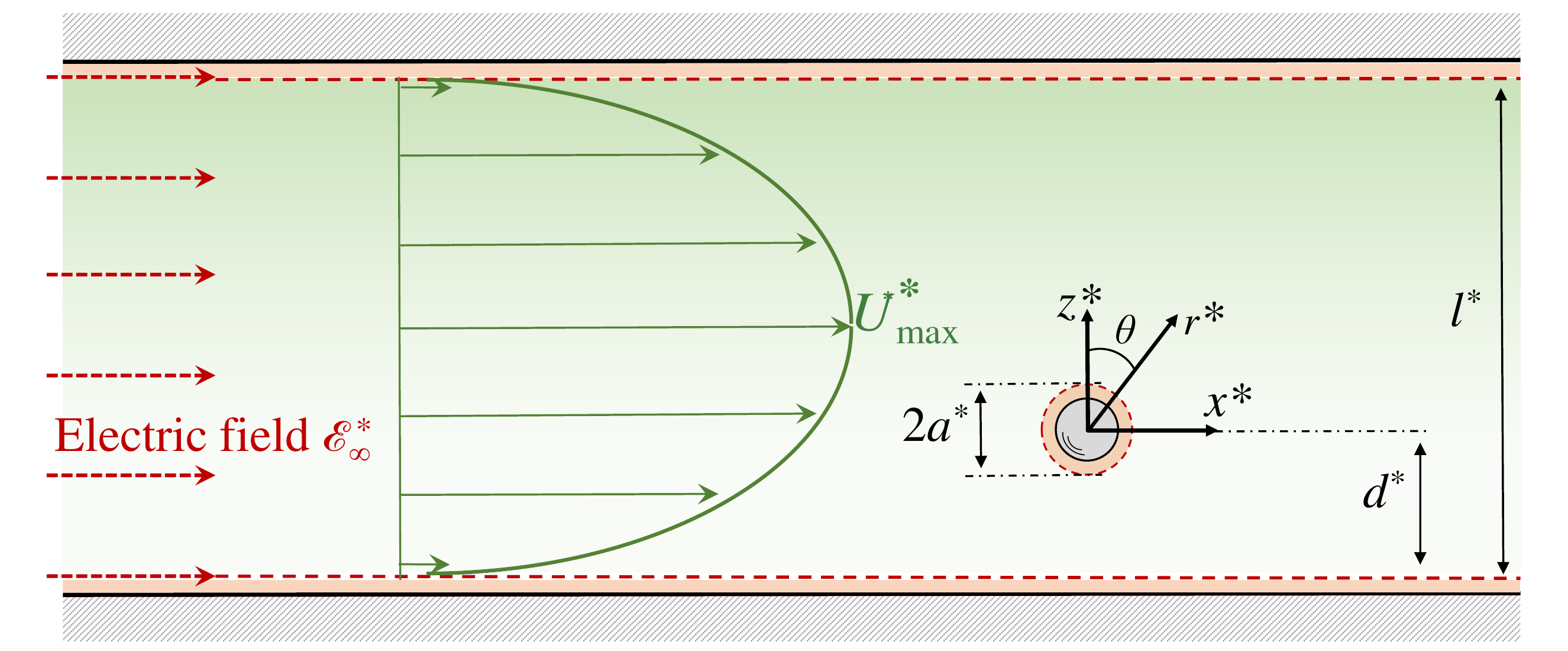}}
 		\caption{Schematic of the problem formulation.The origin of the coordinate system is at the center of the particle. $ y^{*} $ points in the direction of vorticity of the pressure driven flow.
 			The dashed layer around the particle and walls depicts the outer edge of the double layer. Here, * denotes the dimensional variables.}
 		\label{fig:schematic}
 	\end{figure}

 	Figure \ref{fig:schematic} shows the schematic of a neutrally buoyant electrophoretic sphere suspended in a pressure driven flow of SOF, subjected to a parallel electric field ($ \mathcal{E}_\infty^* $). The fluid moves with a maximum velocity of $ U_{max}^{*} $ at the centerline.
 	The frame of reference is placed at the center of the particle.
 	Typical PDMS micro-channel walls and polystyrene particles usually acquire a surface charge (formally quantified as zeta-potential $ \zeta^{*} $) when in contact with an electrolytic solution \citep{zeta_sze2003}, which gives rise to an electrical double layer (EDL) \citep{anderson1989}. 
 	In comparison to the channel and particle size, the EDL can be assumed to be asymptotically thin \textit{i.e.,} $ \lambda^{*}_{D}/a^{*} \ll1 $ (where $\lambda^{*}_{D}$ is the EDL thickness). 
 	Under the influence of an external electric field, the ions inside the EDL \textit{slip} in the tangential direction, generating an electro-osmotic slip. 
 	For low zeta potentials ($ \zeta_{p}^{*} \sim 25 $ mV), the slip at the particle surface can be represented by Smoluchowski's relation \citep{keh1985,anderson1989}.
 	Recently, \cite{ghosh2016electroosmosis} showed that the surface slip in the SOF does not get modified for a surface with constant zeta potential. Hence we employ the Smoluchowski slip in this work.

 	\subsection{Governing equations}
 	We represent the system in terms of dimensionless variables, where the coordinates,  electrostatic potential, velocity and pressure are non-dimensionalized using $ a^{*} $, $ |\mathcal{E}_{\infty}^{*}|a^{*} $, $ \kappa U_{max}^{*} $ and $ \mu^{*} \kappa U_{max}^{*}/a^{*} $, respectively ($ \mu^{*} $ being the dynamic viscosity). 
 	\textcolor{black}{Here $ \kappa $ is the ratio of particle radius ($ a^{*} $) to the channel width ($ l^{*} $). To obtain analytical results, we will later restrict our analysis to small particles ($ \kappa \ll 1 $), which is consistent with the experimental conditions.}
 	
 	The potential ($ \phi $) in the electro-neutral bulk region is governed by the Laplace equation
 	\begin{equation}
 	\nabla^{2} \phi=0
 	\end{equation}
 	The boundary conditions at the particle and wall \textit{surface} (i.e. slip plane of the EDL) are governed by the no-flux condition
 	\begin{equation}
 	{{\IB{e}}_r} \bcdot {\bnabla }\phi=0 \mbox{ at\ } r=1 \quad \mbox{ and\ } \quad
 	{{\IB{e}}_z} \bcdot {\bnabla }\phi=0 \mbox{ at\ } z=-s/\kappa \; \& \; z=(1-s)/\kappa.
 	\label{act_BC}
 	\end{equation}
 	where $ s \equiv d^{*} / l^{*} $ (as defined in fig. \ref{fig:schematic}). The potential is undisturbed ($ \phi_{\infty} \sim -x $) far away from the particle:
 	\begin{equation}
 	\phi  \to {\phi _\infty }
 	\quad \mbox{as\ }\quad r\rightarrow \infty.
 	\end{equation}
 	
 	\vspace{2mm}
 	
 	The hydrodynamics is governed by the continuity and the Navier-Stokes equations:
 		\begin{equation}
 		\bnabla \bcdot \IB{U}=0  \quad \mbox{and\ } \quad {{\mathop{\rm \Rey}\nolimits} _p}\left( { \boldsymbol{U} \bcdot \bnabla \boldsymbol{U}} \right) = \bnabla \bcdot \bten{T},
 		\label{NS_full}
 		\end{equation}
 		where $\Rey_{p}$ is particle Reynolds number defined as: $ {\rho^{*} U_{\max }^{*}\kappa a^{*}}/{\mu^{*}} $ ($ \rho^{*} $ being the fluid density).
 		We model the hydrodynamic stress using the second-order-fluid model \citep{bird1987dynamics}:
 		\begin{equation}
 		\bten{T} = -  P \bten{I} +  \bten{E}^{(1)} + De \, \bten{S}.
 		\end{equation}
 		Here, $ \bten{E}^{(1)} $ represents the rate of strain tensor ($ \bnabla \IB{U} + \bnabla \IB{U} ^{\dagger} $), $ \dagger $ denotes the transpose, $ \bten{S} $ represents the polymeric stress:
 		\begin{equation}
 		\bten{S}= \bten{E}^{(1)} \bcdot \bten{E}^{(1)} + \delta \, \bten{E}^{(2)}.
 		\end{equation}
 		$ \bten{E}^{(2)} $ is the steady-state covariant convected derivative of $ \bten{E}^{(1)} $ (also known as the Rivlin-Ericksen tensor):
 		\begin{equation}
 		\bten{E}^{(2)}= \IB{U} \bcdot \bnabla \bten{E}^{(1)} + \bten{E}^{(1)} \bcdot \bnabla \IB{U}^{\dagger} + \bnabla \IB{U} \bcdot \bten{E}^{(1)},
 		\label{RE_tensor}
 		\end{equation}
 		Here, $ De $ (Deborah number) is the ratio of viscoelastic time scale ($ (\Psi_{1}^{*}+\Psi_{2}^{*})/\mu^{*} $) to the time scale corresponding to the average shear in the background flow ($ l^{*}/U_{max}^{*} $),  $\delta$ is a viscometric parameter, $ \Psi_{1}^{*} $ and $ \Psi_{2}^{*} $ are the steady shear viscometric coefficients \citep{bird1987dynamics}:
 		\begin{equation}
 		De = \frac{(\Psi_{1}^{*} + \Psi_{2}^{*})U_{max}^{*}}{\mu^{*} l^{*}}, \; \delta = \frac{- \Psi_{1}^{*}}{2(\Psi_{1}^{*}+\Psi_{2}^{*})},
 		\;  \Psi_{1}^{*} = \frac{T_{xx}^{*}-T_{zz}^{*}}{ \left( \partial{U_{x}^{*}}/\partial{z^{*}} \right)^{2} }, \;  \Psi_{2}^{*} = \frac{T_{yy}^{*}-T_{zz}^{*}}{\left( \partial{U_{x}^{*}}/\partial{z^{*}} \right)^{2} }.
 		\end{equation}
 	Since the migration time scale is much larger than the convective time scale ($ a^{*}/\kappa U_{max}^{*} $), we analyze the migration in quasi steady-state \citep{becker1996sedimentation,choudhary2019inertial}.
 	Also, we neglect inertial effects (i.e. $ \Rey_{p} \ll De $) which leaves the  polymeric stress as the sole non-linearity in the system.

 	The sphere moves with a translational ($ \IB{U}_{s} $) and rotational ($ \IB{\Omega}_{s} $) velocity, which is to be determined from force and torque balance. Since the domain is infinite in the y-direction, we can consider: $ {\IB{U}_s} = {U_{sx}}{\IB{e}_x} + {U_{sz}}{\IB{e}_z} $  and $ {\IB{\Omega} _s} = {\Omega _{sy}}{\IB{e}_y} $. 
 	To account for the electrokinetic effects in the EDL, we employ the slip-boundary condition derived by \citet{von1903contribution} and impose at the particle surface:
 	\begin{equation}
 	\IB{U} = {\IB{\Omega} _s} \times \IB{r} + H\!a{{\zeta}_p}\IB{\nabla} \phi  
 	\quad \mbox{at\ }\quad r=1.
 	\label{act_BC_particle}
 	\end{equation}
 	Here, $ \zeta_{p} $ is the dimensionless particle surface zeta potential ($ \zeta_{p}^{*}/|\mathcal{E}_{\infty}^{*}|a^{*} $). $ H\!a $ is a Hartmann-type number: $ H\!a=\varepsilon^{*} {\mathcal{E}^{*}}_{\infty}^{2} a^{*}/4\pi \mu^{*} \kappa U_{max}^{*} $, which denotes the ratio of electrical energy density to shear stress. Here $\varepsilon^{*}$ is the electrical permittivity of the medium. 
 	Similarly, the flow satisfies  the Smoluchowski slip condition at the walls
 	\begin{equation}
 	\IB{U} = H\!a{{\zeta}_w}\IB{\bnabla} \phi -  {\IB{U}_s}
 	\quad \mbox{at walls.\ }
 	\label{act_BC_wall}
 	\end{equation}

  	As the particle is freely suspended, the translational ($ \IB{U}_{s} $) and rotational velocity ($ \IB{\Omega}_{s} $) of the particle are obtained using force-free and torque-free conditions: the total force and torque on the particle must be zero. 
 Apart from the hydrodynamic stress, we also have to account for the Maxwell stress in the calculation of total force and torque. Maxwell stress arises due to a lateral asymmetry in the electrostatic potential distribution around the particle due to the presence of the wall.
 The Maxwell stress is defined as
 \begin{equation}\label{Maxwell_stress}
 {\bten{M}} = \bnabla \phi \bnabla \phi  - \frac{1}{2}\left( {\bnabla \phi  \bcdot \bnabla \phi } \right) \bten{I}.
 \end{equation}
 \textcolor{black}{It should be noted that there is no contribution from Maxwell stress to the fluid momentum equation (\ref{NS_full}) because $ \bnabla \bcdot \bten{M} = 0 $ in the electroneutral fluid \citep{yariv2006}.}
 The total force and torque free conditions for particle are given by
 \begin{subequations}\label{FFTF}
 	\begin{gather}
 	\IB{F} = {\IB{F}_H} + {\IB{F}_M} = \int\limits_{{S_p}} {\IB{n} \bcdot {\bten{T}}\mathrm{d}S}   +  4\pi H\!a\int\limits_{{S_p}} {\IB{n} \bcdot {\bten{M}} \, \mathrm{d}S}  = \IB{0}, \\
 	\IB{L} = {\IB{L}_H} + {\IB{L}_M} = \int\limits_{{S_p}} {\IB{n} \times \left( {\IB{n} \bcdot {\bten{T}}} \right)\mathrm{d}S}  + 4\pi H\!a\int\limits_{{S_p}} {\IB{n} \times \left( {\IB{n} \bcdot {\bten{M}}} \,  \right)\mathrm{d}S}  = \IB{0}.
 	\end{gather}
 \end{subequations}	
  	Here subscript $ H $ and $ M $ represent contributions from hydrodynamic and Maxwell stress, respectively.

 	\subsubsection{Undisturbed flow}
 	In the absence of particle, the flow profile for a fully developed pressure-driven second-order fluid flow (in the frame of reference of the particle moving with $ \IB{U}_{s} $) is
 	\begin{equation}
 	{\IB{U}\!\!_\infty} = \left( {\alpha  + \beta z + \gamma {z^2}} \right){\IB{e}_x} + H\!a{\zeta_w}\IB{\bnabla} {\phi _\infty } - {\IB{U}_s}.
 	\label{Vinf}
 	\end{equation}
 	This is identical to that obtained for a Newtonian fluid \citep{choudhary2019inertial}.
 	However, the pressure exhibits a variation in the lateral direction for SOF
 	\begin{equation}
 	{P\!_\infty } = 2 \gamma x + 4 De \, (1+2\delta) (\beta z+\gamma z^{2}) + \mathcal{I},
 	\label{Pinf}
 	\end{equation}
 	where $ \mathcal{I} $ is an integration constant. The constants $ \alpha, \beta$ and $\gamma $ are:
 	\begin{equation}
 	\alpha  = 4s\left( {1 - s} \right)/ \kappa,{\rm{      }}\beta  = 4 \left( {1 - 2s} \right),{\rm{      }}\gamma  =  - 4{\kappa},
 	\label{alpha}
 	\end{equation}
 	where $ \beta $ and $\gamma$ represent the shear and curvature of the background flow.

 	\subsection{Equations governing the disturbance}
 	
 	We analyze the system in terms of disturbance variables (i.e. deviation from the undisturbed flow). For the different variables, these are defined as: potential ($ \psi=\phi-\phi_{\infty} $), velocity ($ \IB{v}=\IB{U}-\IB{U}\!\!_\infty $) and pressure ($ p=P-P_{\infty} $).
 	Since the bulk fluid medium is electroneutral, the disturbance potential is also governed by the Laplace equation
 	\begin{equation}
 	\nabla^{2}\psi=0 .
 	\end{equation}
 	Far away from the particle, $ \psi $ decays to zero. Using the definition of disturbance potential in (\ref{act_BC}):
 	\begin{equation}
 	\IB{e_{r}} \bcdot \bnabla \psi=-\IB{e_{r}} \bcdot \bnabla \phi_{\infty} \, \, \mbox{at\ } r=1
 	\quad \& \quad \IB{e_{z}} \bcdot \bnabla \psi=0 \, \, \mbox{at walls.\ } 
 	\end{equation}
 	The disturbance flow is governed by continuity and Cauchy's momentum equation:
 	\begin{equation}
 	\bnabla \bcdot \IB{v}=0 \quad \mbox{ and\ } \quad \bnabla \bcdot \bten{\sigma}_{H}=\IB{0}.
 	\label{mom}
 	\end{equation}
 	Here $ \IB{\sigma}_{H} $ is the hydrodynamic disturbance stress tensor which contains the Newtonian and polymeric contributions ($ \bten{\sigma}_{H} =  -p\bten{I}+\bten{e}^{(1)} + De\,\bten{s}  $).
 	Here, $ \bten{I} $ is the identity tensor, $ \bten{e}^{(1)} $ is the rate of strain tensor for the disturbance flow ($ \bnabla \IB{v}+\bnabla \IB{v}^{\dagger} $), and $ \bten{s} $ is the disturbance polymeric stress tensor
 	\begin{equation}
 	\bten{s}= \bten{e}^{(1)} \bcdot \bten{e}^{(1)} + \bten{w}^{(1)} + \delta(\bten{e}^{(2)}+\bten{w}^{(2)}),
 	\label{s}
 	\end{equation}
 	{\small
 		\begin{eqnarray}
 		\bten{w}^{(1)} &=& \bten{E}_{\infty}^{(1)} \bcdot \bten{e}^{(1)} + \bten{e}^{(1)} \bcdot \bten{E}_{\infty}^{(1)}, \; \; \;    \bten{e}^{(2)} = \IB{v} \bcdot \bnabla \bten{e}^{(1)} + \bten{e}^{(1)} \bcdot \bnabla \IB{v}^{\dagger} + \bnabla \IB{v} \bcdot \bten{e}^{(1)}, \nonumber \\
 		\bten{w}^{(2)}&=&{\IB{U}\!\!_\infty} \bcdot \bnabla \bten{e}^{(1)} + \bten{e}^{(1)} \bcdot \bnabla {\IB{U}\!\!_\infty}^{\! \! \dagger} + \bnabla {\IB{U}\!\!_\infty} \bcdot \bten{e}^{(1)}
 		+ \, \IB{v} \bcdot \bnabla \bten{E}_{\infty}^{(1)} + \bten{E}_{\infty}^{(1)} \bcdot \bnabla \IB{v}^{\dagger} + \bnabla \IB{v} \bcdot \bten{E}_{\infty}^{(1)}.
 		\end{eqnarray}
 	}
 	Here, $ \bten{E}_{\infty}^{(1)} $  is the rate of strain tensor for the undisturbed flow; $ \bten{e}^{(2)} $ is the disturbance Rivlin-Eriksen (RE) tensor; $ \bten{w}^{(1)} $ is the `interaction tensor' (arising from the interaction between background flow and disturbance flow); and $ \bten{w}^{(2)} $ is the RE interaction tensor. 
 	Using (\ref{act_BC_particle}) and (\ref{act_BC_wall}), we find that the disturbance flow is subject to the following boundary conditions:
 	\begin{subequations}\label{BCvel}
 		\begin{gather}
 		\IB{v} = {\IB{\Omega}_s} \times \IB{r} + H\!a{\zeta_p}\bnabla \left( {\psi  + {\phi _\infty }} \right) - {\IB{U}\!\!_\infty} \quad \mbox{at\ } r=1,\\
 		\IB{v} = H\!a{\zeta_w}\bnabla \psi \quad \mbox{at walls,\ }\\
 		\IB{v} \rightarrow \IB{0} \quad \mbox{as\ } r\rightarrow \infty.
 		\end{gather}
 	\end{subequations}

 The total Maxwell stress can be decomposed into four components as
 \begin{eqnarray}\label{Maxwell}
 	\bten{M} &=& \left( \bnabla \phi_{\infty} \bnabla \phi_{\infty}  - \frac{1}{2}\left( {\bnabla \phi_{\infty}  \bcdot \bnabla \phi_{\infty} } \right) \bten{I}  \right)   +  \left(   \bnabla \psi \bnabla \phi_{\infty}  - \frac{1}{2}\left( {\bnabla \psi  \bcdot \bnabla \phi_{\infty} } \right) \bten{I}  \right) \nonumber \\
 	 && +  \left(   \bnabla \phi_{\infty} \bnabla \psi  - \frac{1}{2}\left( {\bnabla \phi_{\infty}  \bcdot \bnabla \psi} \right) \bten{I}  \right) + \left(   \bnabla \psi \bnabla \psi  - \frac{1}{2}\left( {\bnabla \psi  \bcdot \bnabla \psi} \right) \bten{I}  \right).
 \end{eqnarray}
The first term arises from the undisturbed component, the second and third terms denote the contribution from the interaction between applied ($ \phi_\infty $) and disturbed potential field ($ \psi $), and the fourth term denotes the disturbance Maxwell stress. \cite{yariv2006} showed that the first three components do not result in a net force or torque and the leading order contribution arises from the fourth term{\footnote{\textcolor{black}{Since $ \bten{M} $ is divergence free in the fluid domain, only the terms (in eq.\ref{Maxwell}) decaying as $ O(r^{-2}) $ contribute to the force \citep{guazzelli2011}. Only the disturbance Maxwell stress generates $ O(r^{-2}) $ terms and provides the leading order contribution \citep[p.3]{yariv2006}. }}}. Thus, the disturbance Maxwell stress is denoted as
\begin{equation} \label{distMax}
	\bten{\sigma}_{M} =  \bnabla \psi \bnabla \psi  - \frac{1}{2}\left( {\bnabla \psi  \bcdot \bnabla \psi} \right) \bten{I}.
\end{equation}
Similarly, the undisturbed hydrodynamic field cannot generate a net force or torque on the particle. Hence, the force and torque free conditions are given by
\begin{subequations}\label{FFTF_dist}
	\begin{gather}
	\IB{F} = {\IB{F}_H} + {\IB{F}_M} = \int\limits_{{S_p}} {\IB{n} \bcdot {\bten{\sigma}_{H}}\mathrm{d}S}   +  4\pi H\!a\int\limits_{{S_p}} {\IB{n} \bcdot {\bten{\sigma}_M} \, \mathrm{d}S}  = \IB{0}, \\
	\IB{L} = {\IB{L}_H} + {\IB{L}_M} = \int\limits_{{S_p}} {\IB{n} \times \left( {\IB{n} \bcdot {\bten{\sigma}_{H}}} \right)\mathrm{d}S}  + 4\pi H\!a\int\limits_{{S_p}} {\IB{n} \times \left( {\IB{n} \bcdot {\bten{\sigma}_{M}}} \,  \right)\mathrm{d}S}  = \IB{0}.
	\end{gather}
\end{subequations}

 	\subsection{Migration due to Maxwell stress}
 	Here we evaluate the force and lateral migration arising from the Maxwell stress (\ref{distMax}) alone.
 	Since the potential distribution is not affected by the fluid rheology, the Maxwell force on the particle in a second order fluid is identical to that in a Newtonian fluid \citep{yariv2006}. Following our previous analysis \citep{choudhary2019inertial}, we use the method of reflections to determine the potential distribution in the particle-wall configuration \citep{brenner1962}. 
 	\textcolor{black}{Since our aim is to obtain qualitative insights and analytical results, we restrict the analysis to small particles i.e. $ \kappa \ll 1 $.
 	Assuming the small particle is not too close to the walls (i.e. $ s \gg \kappa $),} the disturbance potential is sought as successive reflections: $ 	\psi  = {{}_{(1)}\psi} + {{}_{(2)}\psi} + \dotsm $. Here, $ {}_{(i)}\psi $ represents the $ i^{th} $ reflection, where the odd reflections satisfy boundary conditions at the particle surface and even reflections satisfy the wall boundary conditions.
 	\textcolor{black}{
 		Accounting for each successive pair of reflections increases the accuracy by $ O(\kappa) $ \citep[p.286]{happel2012low}. Since $ \kappa \ll 1 $, we take only the first two reflections of disturbance potential into account.	}
 	Evaluation of $ \IB{F}_{M} $ and $ \IB{L}_{M} $ (detailed in Appendix A.2) yields:
 	 	\begin{equation}
 	{\IB{F}_M} = 4\pi H\!a\left( {\frac{{3\pi }}{{16}}{\kappa ^4}\left(\mathfrak{Z}{\left({4,s} \right) - \mathfrak{Z} \left( {4,1 - s} \right)} \right)} \right) \IB{e}_{z}  +  O\left( {{\kappa ^7}} \right)\; \text{and} \; {\IB{L}_M} = \IB{0}.
 	\label{FMLM}
 	\end{equation}
 	The Maxwell force acts along the cross-stream direction alone and is wall-repulsive in nature; the Maxwell torque is zero. 	The contribution of Maxwell stresses to the migration velocity is
 	\begin{equation}
 	U_{mig}^{M} \approx {\frac{4\pi H\!a}{6 \pi (1+O(\kappa))}} \left( {\frac{{3\pi }}{{16}}{\kappa ^4}\left(\mathfrak{Z}{\left({4,s} \right) - \mathfrak{Z} \left( {4,1 - s} \right)} \right)} \right)  .
 	\label{MaxVel}
 	\end{equation}
 	Here, superscript $ M $ shows. that the contribution is from Maxwell stress, $ \mathfrak{Z} $ is the generalized Riemann zeta function, and $ 6 \pi (1+O(\kappa)) $ is the viscous resistance which includes wall correction at the first order \citep[p.246]{brenner1961slow}\footnote{\textcolor{black}{The details of evaluation through the method of reflections are provided in the supplementary material.}}. 
 	Later we show that $ U_{mig}^{M} $ decays rapidly away from the walls and is negligible in the bulk of the channel, where migration due to hydrodynamic stress dominates.

 	
 	
 	\subsection{Migration due to hydrodynamic stress}

 	\subsubsection{Perturbation expansion}
 	We find the migration at $ O(De) $ using a regular perturbation expansion. For small values of $ De $, the disturbance field variables are expanded as:
 	\begin{equation}
 	\xi=\xi_{(0)} + De \; \xi_{(1)} + \cdots.
 	\label{pert}
 	\end{equation}
 	Here, $ \xi $ is a generic field variable which represents velocity ($ \IB{v} $), pressure ($ p $), translational ($ \IB{U}_{s} $) and angular velocity ($ \IB{\Omega}_{s} $). Since the electrostatic potential is decoupled from the hydrodynamics, the variable $ \psi $ is not expanded as in (\ref{pert}). 
 	We substitute (\ref{pert}) in the equations governing the hydrodynamics (\ref{mom}-\ref{BCvel}). We obtain the problem at $ O(1) $ as
 	\begin{equation}
 	\left. \begin{array}{l}
 	\qquad \; \quad \bnabla  \bcdot {\IB{v}_{(0)}} = 0, \\
 	{\nabla ^2}{\IB{v}_{(0)}} - \bnabla {p_{(0)}} = \IB{0},\\
 	\qquad \qquad \quad{\IB{v}_{(0)}} = {\IB{\Omega} _s}_{(0)} \times \IB{r} + H\!a{\zeta_p}\bnabla \left( {\psi  + {\phi _\infty }} \right) - {\IB{U}\!\!_\infty} _{(0)} \quad \mbox{at\ } r = 1,\\
 	\qquad \qquad \quad{\IB{v}_{(0)}} = H\!a{\zeta_w}\bnabla \psi \quad \mbox{at walls},\\
 	\qquad \qquad \quad{\IB{v}_{(0)}} \to \IB{0} \quad \mbox{as\ } r \to \infty,
 	\end{array} \right\}
 	\label{Order0}
 	\end{equation}
 	and at $ O(De) $ as:
 	\begin{equation}
 	\left. \begin{array}{l}
 	\qquad \; \quad \bnabla  \bcdot {\IB{v}_{(1)}} = 0, \\
 	{\nabla ^2}{\IB{v}_{(1)}} - \bnabla {p_{(1)}} =  - \bnabla \bcdot \bten{s}_{(0)},\\
 	\qquad \qquad \quad {\IB{v}_{(1)}} = {\IB{U}_{s}}_{(1)} + {\IB{\Omega}_s}_{(1)} \times \IB{r}\quad \mbox{at\ } r = 1,\\
 	\qquad \qquad \quad{\IB{v}_{(1)}} = \IB{0} \quad \mbox{at walls},\\
 	\qquad \qquad \quad{\IB{v}_{(1)}} \to \IB{0} \quad \mbox{as\ } r \to \infty.
 	\end{array} \right\}
 	\label{Order1}
 	\end{equation}
 	Here,  $ \bten{s}_{(0)} $ is the disturbance polymeric stress of $ O(De^{0}) $ field.
 	The angular ($ \IB{\Omega}_{s} $) and translational velocity ($ \IB{U_{s}} $) are evaluated by using (\ref{FFTF_dist}), at each order.

 	\subsubsection{Reciprocal theorem}
 	
 	The symmetry of Stokes flow implies that the $ O(1) $ field cannot produce a lateral lift. Therefore, the lift must arise from $ O(De) $ field. 
 	The reciprocal theorem \citep{ho1976migration} allows us to determine the migration associated with $ \IB{v}_{(1)} $, without having to solve for $ O(De) $ field. 
 	The reciprocal theorem relates the properties of an unknown Stokes flow to a known test flow field, provided both flow fields correspond to the same geometry.
 	The test field ($ \IB{u}^{t},\,p^{t} $) is chosen to be that generated by a moving sphere in the positive z-direction (towards the upper wall) with unit velocity in a quiescent Newtonian fluid (Refer Appendix B for the details of test field). 
 	Following \cite[p.789]{ho1976migration}, we obtain the lateral migration velocity at $ O(De) $ as:
 	\begin{equation}
De (U_{s (1)})_{z}=	U_{mig}^{H}=-{\frac{1}{6 \pi (1+O(\kappa))}}De \; \int_{V_{f}} \bten{s}_{(0)} \colon \bnabla \IB{u}^{t} \: \rm{d}V.
 	\label{mig}
 	\end{equation}
 	Here, superscript $ H $ denotes the hydrodynamic contribution and $ 6 \pi (1+O(\kappa)) $ is the viscous resistance including wall correction at the first order. The details of derivation are shown in the supplementary material.

 	\subsubsection{Evaluation of $ \IB{v}_{(0)} $}
 	Following the method used for the evaluation of potential ($ \psi $), we use the method of reflections to find the velocity field which satisfies the boundary conditions at the particle and wall surface with sufficient accuracy \citep{brenner1962}. \textcolor{black}{As in \S 3.3, we restrict the analysis to small particles i.e. $ \kappa \ll 1 $. This condition is in addition to $ De \ll 1 $, which was used to expand the field variables in (\ref{pert}).}
 	We seek the disturbance ($ \IB{v}_{(0)}, p_{(0)} $) as
 	\begin{equation}
 	{\IB{v}_{\left( 0 \right)}} = {}_{(1)}\IB{v}_{\left( 0 \right)} + {}_{(2)}\IB{v}_{\left( 0 \right)} +  \cdots, \quad 
 	{p_{\left( 0 \right)}} = {}_{(1)}p_{\left( 0 \right)} + {}_{(2)}p_{\left( 0 \right)}  +  \cdots.
 	\label{3.19}
 	\end{equation}
 	Here $ {}_{(i)}\IB{v}_{\left( 0 \right)} $ represents the $ i $\textsuperscript{th} reflection of the creeping flow disturbance velocity. Since $ \kappa \ll 1 $, we take the first two reflections into account.
 	We substitute the above expansion in the equations governing the creeping flow hydrodynamics (\ref{Order0}) and obtain the following set of problems:
 	\begin{equation}
 	\left. \begin{array}{l}
 	\nabla  \cdot {}_{(1)}\IB{v}_{\left( 0 \right)} = 0, \quad {\nabla ^2} {}_{(1)}\IB{v}_{\left( 0 \right)} - \nabla {}_{(1)}p_{\left( 0 \right)} = \IB{0},\\ [1 pt]
 	\; \;\, \quad {}_{(1)}\IB{v}_{\left( 0 \right)} = U_{sx\, \left( 0 \right)}{\IB{e}_x} + \Omega _{sy\, \left( 0 \right)}{\IB{e}_y} \times \IB{r} + H\!a{\zeta_p}(\nabla {{}_{(1)}\psi } + \nabla {{}_{(2)}\psi})\\
 	\; \, \quad \qquad \qquad - \left( {\alpha  + \beta z + \gamma {z^2} + H\!a({\zeta_p} - {\zeta_w})} \right){\IB{e}_x} \quad \mbox{at\ } r = 1,\\
 	\; \; \quad {}_{(1)}\IB{v}_{\left( 0 \right)} \to \IB{0}\qquad \mbox{at\ }r \to \infty .
 	\end{array} \right\}
 	\label{3.20}
 	\end{equation}
 	\begin{equation}
 	\left. \begin{array}{l}
 	\nabla  \cdot {}_{(2)}\IB{v}_{\left( 0 \right)} = 0,\quad {\nabla ^2} {}_{(2)}\IB{v}_{\left( 0 \right)} - \nabla {}_{(2)}p_{\left( 0 \right)} = \IB{0},\\ [1 pt]
 	\; \; \, \quad {}_{(2)}\IB{v}_{\left( 0 \right)} = H\!a{\zeta_w}\left( {\nabla {{}_{(1)}\psi} + \nabla {{}_{(2)}\psi}} \right) - {}_{(1)}\IB{v}_{\left( 0 \right)}\quad \mbox{at the walls,\ }\\
 	\;  \; \, \quad {}_{(2)}\IB{v}_{\left( 0 \right)} \to \IB{0} \quad \mbox{at\ } r \to \infty .
 	\end{array} \right\}
 	\label{3.21}
 	\end{equation}

 	\vspace{1mm}
 	The first reflection ($ {}_{(1)}\IB{v}_{(0)}, \, {}_{(1)}p_{(0)}$) is found by employing Lamb's general solution \citep{lamb,happel2012low}: 
 	\begin{eqnarray}
 	{}_{(1)}\IB{v}_{\left( 0 \right)} &=&  {A_1}\left( {{\IB{e}_x} + \frac{{x\IB{r}}}{{{r^2}}}} \right)\frac{1}{r} + {B_1}\left( { - {\IB{e}_x} + \frac{{3x\IB{r}}}{{{r^2}}}} \right)\frac{1}{{{r^3}}} + {C_1}\left( {\frac{{z{\IB{e}_x}}}{{{r^3}}} - \frac{{x{\IB{e}_z}}}{{{r^3}}}} \right) + {D_1}{\frac{{zx\IB{r}}}{{{r^5}}}}  \nonumber\\
 	&& +  E_1 \left(z \IB{e}_{x} + x \IB{e}_{z} - \frac{5 x z \IB{r}}{r^{2}}\right) \frac{1}{r^{5}}  +\, F_1 \left(\IB{e}_{x} - \frac{2z^{2}\IB{e}_{x} + x\IB{r}}{r^{2}} + \frac{2xz\IB{e}_{z}}{r^{2}} \right)\frac{1}{r^{3}}  \nonumber\\
 	&&   + G_1\left(\IB{e}_{x} - \frac{5z^{2}\IB{e}_{x} +10xz\IB{e}_{z} + 13 x\IB{r} }{r^{2}} + \frac{75xz^{2} \IB{r}}{r^{4}} \right)\frac{1}{r^{3}}  \nonumber\\ 
 	&& + \, H_1\left(\IB{e}_{x} - \frac{5z^{2}\IB{e}_{x} +10xz\IB{e}_{z} + 5 x\IB{r} }{r^{2}} + \frac{35xz^{2}\IB{r}}{r^{4}} \right)\frac{1}{r^{5}}. \label{Lamb_vel} \\ \vspace{2mm}
 	{}_{(1)}p_{\left( 0 \right)} &=&  A_{1} \, \frac{3 x}{2 r^{3}}  + D_{1}  \frac{2 x z}{r^{5}}  +  G_{1} \left( \frac{-30 x}{r^{5}} + \frac{150x z^{2}}{r^{7}} \right) .
 	\label{Lamb_pres}
 	\end{eqnarray}
 	\normalsize
 	Here, the coefficients are defined as:
 	\small
 	\begin{eqnarray}
 	&& A_1=\frac{3}{4} \left( U_{s\,x\,(0)}-\alpha- \frac{\gamma}{3} - H\!a (\zeta_{p}-\zeta_{w})\right) , \; B_1=-\frac{1}{4} \left( U_{s\,x\,(0)}-\alpha- \frac{3\gamma}{5} - H\!a (\zeta_{p}-\zeta_{w}) \right)  + \frac{H\!a \zeta_{p}}{2},\nonumber  \\ 
 	\;   &&C_1=\Omega_{s\,y\,(0)}-\frac{\beta}{2}, 
 	D_1=-\frac{5\beta}{2},  \;  E_1=-\frac{\beta}{2} ,\; F_1=\frac{\gamma}{3}, \;  G_1= -\frac{7 \gamma}{120} ,\; H_1=\frac{\gamma}{8}  .
 	\label{coeff}
 	\end{eqnarray}
 	\normalsize
 		The terms multiplying the coefficients $ A_1 $, $ B_1 $, $ C_1 $, $ D_1 $, $ E_1 $ represent the  force-monopole (stokeslet), source-dipole, antisymmetric dipole (rotlet),  symmetric dipole (stresslet), and octupole singularities, respectively  \citep{kim2013,guazzelli2011}. 
 		\textcolor{black}{
 			The other disturbances (terms multiplying $ F_{1},\, G_{1},\, H_{1} $) are further singularities in the multipole expansion which arise due to the curvature ($ \gamma $) in the background flow field.
 		}
 
 \textcolor{black}{		
The evaluation of higher order reflections can be found using Faxén transformation \citep{faxen1922} and is identical to our previous analysis \citep{choudhary2019inertial}; the details of evaluation are provided in the supplementary material.
We shall later show (in \S3.4.4) that the higher reflections of the velocity field (i.e. $ {}_{(2)}\IB{v}_{(0)} $, $ {}_{(3)}\IB{v}_{(0)}, \cdots $) are dispensable for the current problem.
}

 	The unknown translational ($ {{U}_{s \, x}}_{(0)} $) and rotational ($ {{\Omega}_{s\, y}}_{(0)} $) velocities can be evaluated using the force-free and torque free conditions (\ref{FFTF_dist}) for first three reflections\footnote{Since the second reflection does not contribute to the hydrodynamic drag and torque \citep{happel2012low,kim2013}, we evaluate the third reflection to estimate the wall corrections.}:
 	\begin{subequations} \label{UOmega}
 		\begin{gather}
 		U_{{s\,x}{(0)}} = \alpha+\gamma/3 + H\!a (\zeta_{p}-\zeta_{w}) + O(\kappa^{2}) \\
 		\Omega_{{s\,y}{(0)} } = \beta/2 + O(\kappa^{3}).
 		\end{gather}
 	\end{subequations}
 	\textcolor{black}{
 		The $ O(\kappa^{2}) $ and $ O(\kappa^{3}) $ corrections are contributions from the wall reflections of the disturbance velocity, which arises when the third reflection is taken into account. 
 		The details of the derivation of (\ref{UOmega}) can be found in the supplementary material.
 		Substitution of (\ref{UOmega}) in (\ref{coeff}) yields $ A_{1} \sim O(\kappa^{2}), \; B_{1} \sim H\!a \zeta_{p} O(1) , \; \mbox{and\ } C_{1} \sim  O(\kappa^{3}) $.
 		Since $ \kappa \ll 1 $, we neglect the $ O(\kappa^{2}) $ and $ O(\kappa^{3}) $ wall corrections, and thus use
 		\begin{equation}\label{Usx0}
 			 A_{1}=C_{1}=0 , \; U_{s\,x\,(0)}=\alpha + \gamma/3 + H\!a (\zeta_{p}-\zeta_{w}), \;  \Omega_{{s\,y}{(0)} } = \beta/2 .
 		\end{equation}
 		This yields the background flow velocity in the frame of reference of $ \IB{U}_{s\,(0)} $ (\ref{Vinf})
 		\begin{equation}\label{Vinf0}
 			\IB{U}_{\infty \, (0)} = \left(\beta z + \gamma z^{2} - H\!a \zeta_{p} - \frac{\gamma}{3}\right) \IB{e}_{x}.
 		\end{equation}
 	}
 \vspace{1mm}
 
 		\textcolor{black}{
 	The above estimates of $ A_{1} $, $ C_{1} $ and $ U_{s\,x(0)} $ in (\ref{Usx0}) when substituted in (\ref{Lamb_vel}) provide an insight into leading order velocity disturbances:
 	(i) stresslet ($ \sim O(1/r^{2}) $ arising from the resistance to strain), (ii) source-dipole field ($ \sim O(1/r^{3}) $ arising from the electrophoretic slip), and (iii) octupole field ($ \sim O(1/r^{4}) $ also arising from the resistance to strain). The other disturbances (multiplying $ F_{1},G_{1},H_{1} $), arising due to curvature in the background flow are of $ O(\kappa) $ and are accounted in the evaluation of (\ref{mig}). 
 	}


 	\subsubsection{Evaluation of volume integral}
 	To estimate the migration velocity using (\ref{mig}), we substitute the velocity field $ \IB{v}_{(0)} $ into the integrand.
 	We analyze the intregral for computational convenience and divide the fluid domain ($ V_{f} $) into two asymptotic sub-domains \citep{ho1974,ho1976migration}:
 	\begin{equation}
 	{V_1} = \left\{ {\IB{r} : 1 \le |r| \le \rho } \right\},\, {V_2} = \left\{ {\tilde{\IB{r}} : \kappa \rho  \le |\tilde{r}| < \infty,  -s\le \tilde{z} \le 1-s  } \right\}.
 	\end{equation}
 	Here, the intermediate radius $\rho$ satisfies: $ 1 \ll \rho\ll 1/\kappa $, $\tilde{r}$ represents the $ $\textit{outer coordinates} ($ \tilde{r} = r \kappa $). The characteristic length scale corresponding to the inner domain ($ V_{1} $) is particle radius ‘$ a^{*} $’. Whereas, channel width ‘$ l^{*} $’ is the characteristic length corresponding to the outer domain ($ V_{2} $).
 		Below, we inspect the nature of integrand in each sub-domain and show that the integration in $ V_{2} $ is not required to attain the leading order migration velocity. 
 		
\textcolor{black}{
\justify
		\textit{Integration in sub-domain $ V_{1} $}:
	 Substituting (\ref{Lamb_vel}) and (\ref{Vinf0}) in (\ref{s}) and performing the integration (\ref{mig}) in $ V_{1} $ (which is essentially a spherical shell, provided the particle is not too close to the wall i.e. $ s \gg \kappa $, where $ \kappa \ll 1 $)\footnote{The analytical integration of the volume integral is performed using Mathematica 12. \textcolor{black}{The code is provided in the supplementary material.}}, we obtain
\begin{eqnarray}
	\frac{6\pi(1 \! + \! O(\kappa)) \, U_{mig}^{H}}{De} &=&  - \int_{V_{1}:r=1}^{r=\rho} \bten{s}_{(0)} \colon \bnabla \IB{u}^{t} \: {\rm{d}V} = - \int_{V_{1}:r=1}^{r = \infty} \bten{s}_{(0)} \colon \bnabla \IB{u}^{t} \: {\rm{d}V} + o(\kappa)  \nonumber \\ 
	&\approx& \frac{10 \pi}{3} \beta \gamma \left ( 1 \! + \! 3 \delta  \right)  \! 
	- \! \frac{3\pi}{2} \beta H\!a \zeta_{p} \left ( 1 \! + \! \delta  \right).
	\label{final}
\end{eqnarray}
The details of $ o(\kappa) $ error incurred while approximating the upper limit of integral to $ r = \infty $ is detailed in Appendix C.
}

 	In view of the definition of $ \beta $ and $ \gamma $, the first term in (\ref{final}) is $ O(\kappa) $, and is identical to that obtained by \cite{ho1976migration}, who reported the viscoelastic migration of a particle in the absence of electrokinetic effects. 
 	The second term is $  O(1) $ and is the dominant contribution to the migration in the current study. It is interesting to note that the first term arises purely from the interaction of disturbance field and the bulk flow (i.e. $ \bten{w}^{(1)}_{(0)}  \mbox{ and\ } \bten{w}^{(2)}_{(0)} $). 
 		However, the second term arises from the interactions among various disturbances (i.e. $ \bten{e}^{(1)}_{(0)} \bcdot \bten{e}^{(1)}_{(0)} \mbox{ and\ } \bten{e}^{(2)}_{(0)} $).
 		This insight and the proportionality to $ \beta H\!a \zeta_{p}  $ (second term) suggests that the $ O(1) $ modification to migration arises from the interaction of stresslet and octupole disturbances (both proportional to $ \beta $) with the electrophoresis induced source dipole disturbance (proportional to $ H\!a \zeta_{p} $).
 	
\vspace{2mm}
\justify
 		 	\textcolor{black}{	\textit{Integration in sub-domain $ V_{2} $}:	}
 	We now find the contribution of the integral in the outer subdomain $ V_{2} $. Since the characteristic length in the outer domain is $ l^{*} $, the disturbances are stretched into the outer coordinates ($ \tilde{r}=r \kappa $). In the outer coordinate representation, the order of magnitude of the leading order disturbance and test field velocity is
 		\begin{subequations}\label{OOM_distOuter}
 			\begin{gather}
 			\tilde{\IB{v}}_{(0)} \sim   H\!a \zeta_{p} O(\kappa^{3})  + \beta \, O(\kappa^{2}) + \beta \, O(\kappa^{4}) + \gamma \, O(\kappa^{3})  + \gamma \, O(\kappa^{5}), \label{OOM_velOuter}\\
 			\tilde{\IB{u}}^{t} \sim  O(\kappa) + O(\kappa^{3}). \label{OOM_testOuter}
 			\end{gather}
 		\end{subequations}
 		In the outer coordinates, the undisturbed background flow velocity is
 		\begin{equation}
 		{\tilde{\IB{U}}\!\!_{\infty\, (0)}} \sim  \beta \, O(1/\kappa) + \gamma \, O(1/\kappa^{2}) + H\!a \zeta_{p} O(1) + \gamma \, O(1).
 		\label{OOM_VinfOuter}
 		\end{equation}
 		We substitute (\ref{OOM_velOuter}) and (\ref{OOM_VinfOuter}) in (\ref{s}) and obtain $ \tilde{\bten{s}}_{(0)} $.

 		The volume integral (\ref{mig}), in $ V_{2} $ subdomain, is represented in outer coordinates as 
 		\begin{equation}
 	\frac{6\pi(1 \! + \! O(\kappa)) \, U_{mig}^{H}}{De} = -	\int_{V_{2}}\tilde{\bten{s}}_{(0)} \colon \left( \kappa \tilde{\bnabla} \tilde{\IB{u}}^{t}\right) \kappa^{-3} {\rm{d}} \tilde{V}.
 		\end{equation}
 		Upon simplification, we find that the integrand is $ O(\kappa^{2}) $. Since $ \kappa \ll 1 $, the dominant contribution can be concluded to arise from the inner integral (i.e. $ O(1) $) and the contribution from the outer integral can be neglected.
 	In Appendix D.1, the expression for leading order migration velocity (eq. \ref{final}) is verified using an independent calculation following the theorem of \cite{g3d} for the special case of polymeric fluids ($ \Psi_{2}^{*} = -0.5 \, \Psi_{1}^{*} $).


 	\section{Results and discussion}
 	
 	The expression for the migration velocity (\ref{final}) depicts that the direction of lateral migration is determined by the magnitude and sign of the viscometric parameter $ \delta \mbox{ and\ } De  $. For most viscoelastic fluids $ De $ is positive, and $\delta$ is reported to be between -0.5 and -0.7 \citep{caswell1962creeping,leal_1975,koch2006stress,bird1987dynamics}.
 	Since $ \gamma $ is negative, the first term in (\ref{final}) is strictly positive below the centerline ($ \beta>0 $) and negative above it ($ \beta<0 $) \textit{i.e.} the pure elastic stresses push the particle towards the channel axis at $ O(\kappa) $. 
 	The second term is $  O(1) $; its proportionality to $ H\!a \zeta_{p} $ indicates that this dominant contribution originates due to the electrophoretic slip \textit{i.e.,} incorporation of electrophoresis enhances the migration by $ O(\kappa^{-1}) $.
 	For $ -0.7<\delta<-0.5 $, a particle with positive $ \zeta_{p} $ will migrate towards the walls (at $ O(1) $), when the electric field is applied in the direction of the flow. The reverse would hold for a particle with negative $ \zeta_{p} $ or if the electric field is applied in the opposite direction. 
 	These findings qualitatively agree with the experimental observations depicted in Fig.\ref{fig:exp}.

 	The enhancement demonstrates that an addition of electrokinetic slip on the particle surface can drastically alter the normal stresses which consequently enhances the particle migration. The slip creates a source-dipole disturbance in the fluid which decays rapidly away from the particle ($ \sim O(1/r^{3}) $). This suggests that the migration occurs from short-ranged interactions, which are unaffected by the boundaries unless the particle is very close to the walls.
 	Furthermore, (\ref{final}) reveals that the parameters such as wall zeta potential ($ \zeta_{w} $) and curvature of the background flow ($ \gamma $) do not affect the migration at the present order of approximation.

 	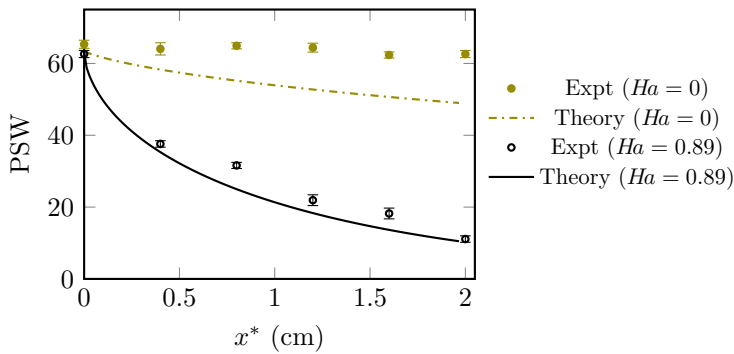
\begin{figure}
 		\centering
 		\begin{tikzpicture}
 		%
 		%
 		%
 		%
 		%
 		%
 		%
 		\begin{axis}[clip=false, width=0.5\textwidth,height=0.38\textwidth,xlabel=$ x^{*}$ (cm), ylabel=PSW, 
 		ylabel near ticks,
 		xmin=0, xmax=2.05 , ymin=0, ymax=75,  thick, legend style={draw=none,at={(1.7,0.3)},anchor=south east},legend style={nodes={scale=0.85, transform shape}
 			,
 			legend columns=1, 
 			,	/tikz/column 2/.style={
 				column sep=5pt,
 			},
 		},]
 		\addplot[olive , mark=*, mark options={scale=0.6}, only marks, error bars/.cd, y dir=both, y explicit] table[x=Len, y=PSW, y error= err] {
 			Len	PSW	err
 			0	65.33	1.1282287
 			0.4	64.05	1.705002444
 			0.8	64.91	0.855004873
 			1.2	64.39	1.239771484
 			1.6	62.35	0.855004873
 			2	62.61	0.989764282
 			
 		};
 		
 		\addplot[olive, dash dot, thick] table[x=x, y=LEAL] {Traj_Aku2.txt};
 		
 		\addplot[black, mark=o, mark options={scale=0.6}, only marks, error bars/.cd, y dir=both, y explicit] table[x=Len, y=PSW, y error= err] {
 			Len	PSW	err
 			0	62.62	1
 			0.4	37.58	0.9
 			0.8	31.6	0.85
 			1.2	21.92	1.5
 			1.6	18.22	1.5
 			2	11.1	0.9
 		};
 		
 		\addplot[black, thick] table[x=x, y=AKU] {Traj_Aku2.txt};
 		\legend{Expt ($ H\!a=0 $), Theory ($ H\!a=0 $) , Expt ($ H\!a=0.89 $), Theory ($ H\!a=0.89 $)}
 		\end{axis}
 		\end{tikzpicture}

 		\caption{Particle stream width (PSW) ($ \mu $m) of negative zeta potential particles suspended in a Poiseuille flow of second order fluid.
 			Parameters: $ De=0.12 $, $ \delta=-0.5 $, $ E_{\infty}^{*}=300 $ V/cm, and $ \zeta_{p}=-2.52 $ ($ \zeta_{p}^{*}=-83 $ mV), $ l^{*}= 66 \, \mu $m, $ \kappa =0.017$.}
 		\label{fig:results}
 	\end{figure}

 	Since the experimental parameters are in the regime of the theoretical analysis ($ De<1 $, $ De \gg Re_{p} $, $ H\!a \zeta_{p} \sim O(1) $, $\kappa \ll 1$), we now quantitatively compare the two.
 	Theoretically, the trajectories of a single particle are calculated using (\ref{final}) and are converted to an equivalent particle stream-width using $ |l^{*}-2(l^{*}-d^{*})| $, for particle focusing at the centerline ($ l^{*} $  being the channel height and $ d^{*} $ is the particle to wall distance)\footnote{In Appendix D.2, the theoretical calculation of trajectory is detailed and are validated with that reported by \cite{ho1976migration} for $ H\!a=0 $.}.
 	We use the relaxation time scale ($ (\Psi_{1}^{*}+\Psi_{2}^{*})/\mu^{*} $) corresponding to $ C^{*}_{PEO}=250 $ ppm as $ 2.75 $ ms \citep{lu2015elasto}. 
 	Figure \ref{fig:results} shows the comparison of particle stream widths for two cases: (i) No electric field (\textit{i.e.,} viscoelastic focusing) and (ii) parallel electric field (\textit{i.e.,} electro-viscoelastic focusing). 
 	For case-(i), the theoretical prediction shows modest particle migration. The experiments show negligible focusing as some particles get trapped near the channel corners, which is one of the limitations of viscoelastic focusing at low flowrates \citep{xuan2017review,tian2019manipulation}. Since our model is 2D, it does not account for the trapping of particles at 3D corners.
 	For case-(ii), the theoretical predictions and experimental results confirm an order of magnitude enhancement in focusing. The trapping of particles at the corners is not seen here as the repulsive Maxwell lift is strongest near the corners \citep{liang2010three}. Thus, the agreement between the experiments and theoretical analysis is improved.

 	\begin{figure}
 		\centering
 		\begin{tikzpicture}
 		\begin{axis}[ ylabel near ticks, xshift=0.66cm,yshift=0.2cm,width=0.26\textwidth,height=0.21\textwidth, extra y ticks= 0,
 		extra y tick labels = ,
 		extra y tick style  = { grid = major },
 		xtick=,
 		xticklabels=,
 		xmin=0, xmax=1 , ymin=-0.00000035 , ymax=0.00000035 , thick, legend style={draw=none,at={(2,1.0)},anchor=south east}]
 		
 		\addplot[line width=0.9pt, black,dashed] table[x=x, y=MM] {Components.txt};
 		\addplot[blue] table[x=x, y=VM] {Components.txt};
 		\end{axis}

 		\begin{axis}[clip=false,
 		width=0.52\textwidth,height=0.40\textwidth, extra y ticks= 0,	ylabel shift= -8 pt,
 		extra y tick labels = ,
 		extra y tick style  = { grid = major }, 
 		xlabel=$ d^{*}/l^{*} $, ylabel=$ U_{L}^{*} $,%
 		, xmin=0, xmax=1, ymin=-0.0000048 , ymax=0.0000048, thick
 		,	legend style={draw=none,at={(1,0.54)},anchor=south east},	legend style={nodes={scale=0.8, transform shape}}]
 		\addplot[line width=1.2pt, red,dash dot] table[x=x, y=EVM] {Components.txt};
 		\addplot[clip mode=individual, 
 		line width=1.1pt, black,dashed] table[x=x, y=MM] {Components.txt};
 		\addplot[blue] table[x=x, y=VM] {Components.txt};
 		\legend{EVM,MM,VM}
 		\end{axis}
 		\end{tikzpicture}
 		\caption{Comparison of various components of the migration velocity (m/s) for $ \kappa=0.017 $; MM (Maxwell migration), VM (Viscoelastic migration), and EVM (Electro-Viscoelastic migration). Here EVM corresponds to Fig.\ref{fig:exp}(b). The inset shows the comparison of MM and VM. \textcolor{black}{Parameters: $ De=0.12 $, $ \delta=-0.5 $, $ E_{\infty}^{*}=300 $ V/cm, and $ \zeta_{p}=-2.52 $ ($ \zeta_{p}^{*}=-83 $ mV), $ l^{*}= 66 \, \mu $m, $ \kappa =0.017$.}}
 		\label{fig:compare}
 	\end{figure}
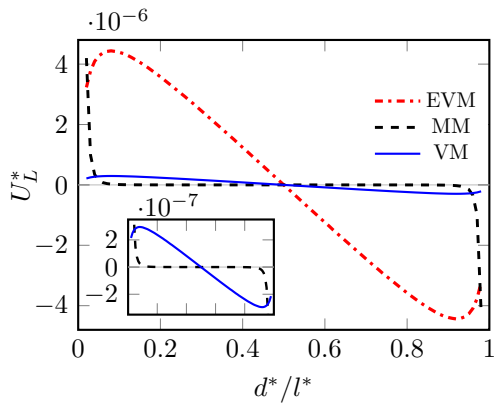
 	
 	Our mathematical framework reveals that the migration in Fig.\ref{fig:exp}(b,c) arises from: (i) Maxwell stresses (eq.\ref{MaxVel}), (ii) pure viscoelastic stresses (eq.\ref{final}-term1), and (iii) electrophoresis enhanced viscoelastic stresses (eq.\ref{final}-term2). A comparison of these migration velocities, namely, Maxwell migration (MM), viscoelastic migration (VM) and electro-viscoelastic migration (EVM) is shown in Fig.\ref{fig:compare}. 
 	The curves correspond to the case in Fig.\ref{fig:exp}(b): a particle with negative zeta potential and electric field applied parallel to the flow.
 	The migration due to decoupled Maxwell stresses (MM) is dominant near the walls and decays rapidly away from it. Although this wall-repulsive force is negligible in the bulk of the channel (for $ \kappa \ll 1 $), it helps eliminate the corner equilibrium positions \citep{liang2010three}.
 	The other components VM and EVM act towards the center as depicted by the negative gradient at the central equilibrium position. 
 	For conditions corresponding to Fig.\ref{fig:exp}(c), where flow and electric field are opposite to each other ($ H\!a <0 $), the EVM curve would reverse, as we can see from (\ref{final}). 
 	As a consequence, the central equilibrium position would become unstable, resulting in particle migration towards the walls.

 	\section{Conclusions}
 	Our experimental and theoretical analysis depict an enhancement of $ O(\kappa^{-1}) $ in migration of a particle in viscoelastic flow, when subjected to a parallel electric field. 
 	Here $ \kappa $ is the ratio of particle radius to channel size.
 	Using a second-order fluid model, the theoretical analysis reveals the contribution from different underlying stresses around the particle (see eq.\ref{MaxVel} and eq.\ref{final}). These constitute an approximate expression for the migration velocity which agrees well with the experimental results. 
 	The analytical results further reveal that the particle zeta potential and background shear determine the equilibrium positions. Wall zeta potential, Maxwell stresses, and curvature of the background flow do not affect the migration at the leading order.
 	
 	\cite{karnis1966particle,gauthier1971particle} suggested that the normal stresses in the undisturbed flow (i.e. tensioned streamlines) are responsible for the  migration of the neutral sphere towards the channel center. 
 	Later, \cite{ho1976migration} reported that this explanation might not be entirely accurate because of the presence of strong disturbance flow around the sphere. 
 	However, the disturbance due to a freely suspended particle in the shear flow does not exhibit a stagnation point, and thus the profile is still predominantly a shear flow \citep{ho1976migration, guazzelli2011}. 
 	Thus the notion of streamlined tension is not altogether lost and can be applied for a neutral particle.
 	For the case of an electrophoretic particle, the above explanation may not hold as the characteristic disturbance field (being a source dipole) exhibits stagnation points, which should be responsible for the direction of migration addressed in this work. This intriguing feature entails a detailed, perhaps numerical, description of pressure and stress fields around the electrophoretic particle to test this speculation, and warrants further investigation.

 	The findings of the current work have a direct practical relevance to applications related to Lab-on-a-chip devices used to separate cells in biological fluids which are known to exhibit non-Newtonian behavior \citep{yuan2018reviewVisco,stoecklein2018nonlinear}. The current work addresses the particle migration in dilute polymer concentration regime. 
 	For higher polymer concentrations, the direction of migration is reported to be opposite to that of the current work and the explanation for this is currently unknown \citep{xuan2018}. 
 	Several studies \citep{rodd2005inertio,yang2011sheathless,kim2012lateral} have reported that flows consisting semi-dilute polymer concentration regime may exhibit shear-thinning behavior.
 	In such cases, constitutive models like FENE-P and Giesekus fluid can be implemented numerically to accurately predict the particle migration. In our future studies we intend to explore these areas.\\ \\

 	The financial support from Indian Ministry of Human Resource Development is gratefully acknowledged. \\ \\
 	
 	\textbf{Declaration of Interests}\\
 	The authors report no conflict of interest.\\ \\
 	
 	\appendix
 	\section{}\label{appA} 
 	\subsection{Evaluation of $ \psi $ using the method of  reflections}

 	To satisfy the boundary conditions at both the particle and wall surface, we seek the disturbance potential as a sum of reflections:
 	\begin{equation}
 	\psi  = {{}_{(1)}\psi} + {{}_{(2)}\psi} + \dotsm.
 	\label{1.1}
 	\end{equation}
 	Here, $ {}_{(i)}\psi $ represents the i\textsuperscript{th} reflection. 
 	The odd reflections satisfy boundary conditions at the particle surface and even reflections satisfy wall boundary conditions.
 	Substituting the above equation in Laplace equation ($ \nabla^{2}{\psi}=0 $), we obtain for the first reflection:
 	\begin{equation}
 	\left. \begin{array}{l}
 	\; \, \quad {\nabla ^2}{{}_{(1)}\psi} = 0,\\
 	{\IB{e}_r} \cdot \nabla {{}_{(1)}\psi} =  - {\IB{e}_r} \cdot \nabla {\phi _\infty }\quad \mbox{at\ } r=1,\\
 	\; \;\, \qquad {{}_{(1)}\psi} \to 0 \quad \mbox{as\ } r \to \infty,
 	\end{array} \right\}
 	\label{1.2}
 	\end{equation}
 	and for the second reflection:
 	\begin{equation}
 	\left. \begin{array}{l}
 	\; \,\quad {\nabla ^2}{{}_{(2)}\psi} = 0,\\
 	{\IB{e}_z} \cdot \nabla {{}_{(2)}\psi} =  - {\IB{e}_z} \cdot \nabla {}_{(1)}\psi \quad \mbox{at walls.\ }
 	\end{array} \right\}
 	\label{1.3}
 	\end{equation}
 	
 	\textit{ Solution to} $ {{}_{(1)}}\psi $: Equation (\ref{1.2}) suggests that $ {{}_{(1)}}\psi $ is a harmonic function which decays as $ r \to \infty $. The boundary condition suggests that it must be linear in the ‘driving force’: $ \nabla {\phi _\infty } (=-{\IB{e}_x}) $. Therefore, the solution is sought in terms of spherical solid harmonics \citep{guazzelli2011} as:
 	\begin{equation}
 	{{}_{(1)}\psi} = \frac{1}{2}\frac{\IB{r}}{{{r^3}}}.\nabla {\phi _\infty } =  - \frac{1}{2}\frac{x}{{{r^3}}}.
 	\label{1.4}
 	\end{equation}
 	
 	\textit{ Solution to} $ {{}_{(2)}}\psi $: To evaluate $ {{}_{(2)}}\psi $, we adopt the approach devised by \citet{faxen1922} in the context of bounded viscous flows. Using this, the disturbance around the particle is expressed in an integral form which satisfies the boundary conditions at the walls. The first reflection is characterized by particle scale ($ a^{*} $). The second reflection $ {{}_{(2)}}\psi $ is characterized by a length scale $ l^{*} $. Therefore, the coordinates for second reflection are stretched, and are termed as `outer' coordinates. These outer coordinates (denoted by tilde) are defined as:
 	\begin{equation}
 	\tilde{r} = r\kappa \quad \tilde{x} = x\kappa \quad \tilde{y} = y\kappa \quad \tilde{z} = z\kappa.
 	\label{1.5}
 	\end{equation}
 	
 	Since $ {{}_{(2)}}\psi $ depends on $ {{}_{(1)}}\psi $ through the wall boundary condition in (\ref{1.3}), both reflections must be represented in the coordinates of same scale. The first reflection of potential disturbance in the outer coordinates ($ {{}_{(1)}}\tilde{\psi} $) is represented as:
 	\begin{equation}
 	{{}_{(1)}\tilde \psi} =  - \frac{1}{2}\frac{\tilde{x}}{{{\tilde{r}^3}}}{\kappa ^2}.
 	\label{1.6}
 	\end{equation}
 	\citet{faxen1922} represented the fundamental solution of Laplace equation (in Cartesian space) in the terms of Fourier integrals as:
 	\begin{equation}
 	\frac{1}{\tilde{r}} = \frac{1}{{2\pi }}\int\limits_{ - \infty }^{ + \infty } {\int\limits_{ - \infty }^{ + \infty } {{\mathrm{e}^{\left( {\mathrm{i}\Theta  - \frac{{\lambda \left| \tilde{z} \right|}}{2}} \right)}}\frac{{ \mathrm{d}\xi \mathrm{d}\eta }}{{2\lambda }}} } \mbox{, and\ }
 	\label{1.7}
 	\end{equation}
 	Here, $ \Theta  = (\xi \tilde{x}+ \eta \tilde{y})/2 $ and $ \lambda=(\xi^{2}+\eta^{2})^{1/2} $ where $ \xi $ and $ \eta $ are the variables in Fourier space. We transform the disturbance field (\ref{1.6}) by taking derivatives of the above equation (\ref{1.7}). For example: (\ref{1.6}) contains $ \tilde{x}/\tilde{r}^{3} $, which is expressed using the above transformation as:
 	\begin{equation}
 	\frac{\tilde{x}}{{{\tilde{r}^3}}} =  - \frac{\partial }{{\partial \tilde{x}}}\left( {\frac{1}{\tilde{r}}} \right) = \frac{1}{{2\pi }}\int\limits_{ - \infty }^{ + \infty } {\int\limits_{ - \infty }^{ + \infty } {{\mathrm{e}^{\left( {\mathrm{i}\Theta  - \frac{{\lambda \left| \tilde{z} \right|}}{2}} \right)}}\left( { - \frac{{\mathrm{i}\xi }}{2}} \right)\frac{{ \mathrm{d}\xi \mathrm{d}\eta }}{{2\lambda }}} } .
 	\label{1.9}
 	\end{equation}
 	This yields an integral representation for the first reflection:
 	\begin{equation}
 	{{}_{(1)}\tilde \psi} = \frac{{{\kappa ^2}}}{{2\pi }}\int\limits_{ - \infty }^{ + \infty } {\int\limits_{ - \infty }^{ + \infty } {{\mathrm{e}^{\left( {\mathrm{i}\Theta  - \frac{{\lambda \left| \tilde{z} \right|}}{2}} \right)}} {b_1}\left( {\frac{{\mathrm{i}\xi }}{\lambda }} \right) \mathrm{d}\xi \mathrm{d}\eta } }.
 	\label{1.10}
 	\end{equation}
 	Here, $ b_{1}=1/8 $. The boundary condition expressed in (\ref{1.3}) suggests a form of solution for $ {{}_{(2)}}\tilde{\psi} $ similar to the above equation:
 	\begin{equation}
 	{{}_{(2)}\tilde \psi} = \frac{{{\kappa ^2}}}{{2\pi }}\int\limits_{ - \infty }^{ + \infty } {\int\limits_{ - \infty }^{ + \infty } {{\mathrm{e}^{\mathrm{i}\Theta }}\left( {{\mathrm{e}^{\left( { - \frac{{\lambda \tilde{z}}}{2}} \right)}}{b_2} + {\mathrm{e}^{\left( { + \frac{{\lambda \tilde{z}}}{2}} \right)}}{b_3}} \right)\left( {\frac{{\mathrm{i}\xi }}{\lambda }} \right) \mathrm{d}\xi \mathrm{d}\eta } }.
 	\label{1.11}
 	\end{equation}
 	Substituting (\ref{1.10}) and (\ref{1.11}) in the boundary condition (\ref{1.3}), yields $ b_{2} $ and $ b_{3} $ in terms of $ b_{1} $:
 	\begin{equation}
 	{b_2} = \frac{{{b_1}\left( {1 + {\mathrm{e}^{\left( {1 - s} \right)\lambda }}} \right)}}{{ - 1 + {\mathrm{e}^\lambda }}},\quad {b_3} = \frac{{{b_1}\left( {1 + {\mathrm{e}^{s\lambda }}} \right)}}{{ - 1 + {\mathrm{e}^\lambda }}}.
 	\label{1.12}
 	\end{equation}
 	The next reflection ($ {{}_{(3)}}\psi $) is $ O(\kappa^{3}) $. As we restrict $ \kappa \ll 1 $, the first two reflections capture the leading order contribution to the disturbance potential.

 	\subsection{Evaluation of Maxwell migration}
 	The electric force and torque are expressed as a function of reflections as:
 	\begin{equation}
 	{\IB{F}_M} = 4\pi H\!a\int\limits_{{S_p}} {\IB{n} \bcdot \left( \begin{array}{l}
 		\bnabla \left( {{{}_{(1)}\psi} + {{}_{(2)}\psi}} \right) \bnabla \left( {{{}_{(1)}\psi} + {{}_{(2)}\psi}} \right)\\
 		{\rm{             }} - \frac{1}{2}\left( {\bnabla \left( {{{}_{(1)}\psi} + {{}_{(2)}\psi}} \right) \bcdot \bnabla \left( {{{}_{(1)}\psi} + {{}_{(2)}\psi}} \right)} \right)\mathsfbi I
 		\end{array} \right)\mathrm{d}S}, 
 	\label{FMMor}
 	\end{equation}
 	\begin{equation}
 	{\IB{L}_M} = 4\pi H\!a\int\limits_{{S_p}} {\IB{n} \times \left( {\IB{n} \bcdot \left( \begin{array}{l}
 			\bnabla \left( {{{}_{(1)}\psi} + {{}_{(2)}\psi}} \right) \bnabla \left( {{{}_{(1)}\psi} + {{}_{(2)}\psi}} \right)\\
 			{\rm{             }} - \frac{1}{2}\left( {\bnabla \left( {{{}_{(1)}\psi} + {{}_{(2)}\psi}} \right) \bcdot \bnabla \left( {{{}_{(1)}\psi} + {{}_{(2)}\psi}} \right)} \right)\mathsfbi I
 			\end{array} \right)} \right)\mathrm{d}S}.
 	\label{LMMor}
 	\end{equation}
 	The gradient of first reflection ($ \bnabla {{}_{(1)}\psi} $) at the particle surface is:
 	\begin{equation}
 	{\left. {\bnabla {{}_{(1)}\psi}} \right|_{r = 1}} =  \frac{1}{2} \left\{ {-1+3 \sin ^2\theta  \cos ^2\varphi, \; 3 \sin ^2\theta  \sin \varphi  \cos \varphi , \; 3 \sin \theta \cos \theta  \cos \varphi } \right\}.
 	\label{DelPsi1BC}
 	\end{equation}
 	Here, $ \varphi $ and $ \theta $ denote the azimuthal and polar angle, respectively. As the wall reflection of the potential is represented in the outer scaled coordinates, $ \bnabla {{}_{(2)}\psi} $ at the particle surface ($ r=1 $) is expressed as a Taylor series expansion of $ \kappa \tilde{\bnabla} \tilde{\psi_{2}} $ about the origin ($ \tilde{r} \rightarrow 0 $). Here $ \tilde{\bnabla} $ is the spatial gradient in outer scaled coordinates: $ \left\{ {\frac{\partial}{\partial \tilde{x}},\, \frac{\partial}{\partial \tilde{y}}, \, \frac{\partial}{\partial \tilde{z}}} \right\} $. Since the particle is small with respect to the wall-particle gap ($ \kappa/s \ll 1 $), we retain only the zeroth and first order terms:
 	\begin{equation}
 	{\left. {\bnabla {{}_{(2)}\psi}} \right|_{r = 1}} = \int\limits_0^\infty  {\left[ \begin{array}{l}
 		{\kappa ^3}\left( {{{ - ({b_2} + {b_3}){\lambda ^2}} \mathord{\left/
 					{\vphantom {{ - ({b_2} + {b_3}){\lambda ^2}} 4}} \right.
 					\kern-\nulldelimiterspace} 4}} \right)\left\{ {1, 0, 0} \right\}\\
 		+ {\kappa ^4}\left( {{{({b_2} - {b_3}){\lambda ^3}} \mathord{\left/
 					{\vphantom {{({b_2} - {b_3}){\lambda ^3}} 8}} \right.
 					\kern-\nulldelimiterspace} 8}} \right)\left\{ {1, 0, 1} \right\}
 		\end{array} \right]}  \mathrm{d}\lambda   +  O({\kappa ^5}).
 	\label{DelPsi2BC}
 	\end{equation}
 	Substituting $ {{}_{(1)}\psi} $ $ \& $ $ {{}_{(2)}\psi} $ in both (\ref{FMMor}) $ \& $ (\ref{LMMor}) and using (\ref{DelPsi1BC}) $ \& $ (\ref{DelPsi2BC}), we obtain:
 	\begin{equation}
 	{\IB{F}_M} = 4\pi H\!a\left( {\frac{{3\pi }}{{16}}{\kappa ^4}\left(\mathfrak{Z}{\left({4,s} \right) - \mathfrak{Z} \left( {4,1 - s} \right)} \right)} \right) \IB{e}_{z}  +  O\left( {{\kappa ^7}} \right)\; \text{and} \; {\IB{L}_M} = \IB{0}.
 	\label{FMLMApp}
 	\end{equation}

 	\section{Test field ($ \IB{u}^{t} $)}
 	The test field is chosen to be that generated by a sphere moving in the positive z-direction with a unit velocity in a quiescent Newtonian medium.
 	\begin{equation}
 	\left. \begin{array}{l}
 	\bnabla  \bcdot {\IB{u}^t} = 0, \; \; \; {\nabla ^2}{\IB{u}^t} - \bnabla {p^t} = \IB{0},\\
 	\; \; \,\quad {\IB{u}^t} = {\IB{e}_z} \quad \mbox{at\ } r = 1,\\
 	\; \; \,\quad{\IB{u}^t} = \IB{0} \quad \; \, \mbox{at the walls,\ }\\
 	\; \; \, \quad{\IB{u}^t} \to \IB{0 }\quad  \;  \mbox{at r} \to \infty .
 	\end{array} \right\}
 	\label{Test}
 	\end{equation} 
 	To solve for $ \IB{u}^{t} $, we again make use of the fact that the integration in $ V_{1} $ is sufficient for the leading order solution and solve for the first reflection of $ \IB{u}^{t} $ using Lamb's solution:
 	\begin{equation}
 	{{}_{(1)}\IB{u}^t} = \frac{3}{4}\left( {{\IB{e}_z} + \frac{{z\IB{r}}}{{{r^2}}}} \right)\frac{1}{r} + \frac{1}{4}\left( {{\IB{e}_z} - \frac{{3z\IB{r}}}{{{r^2}}}} \right)\frac{1}{{{r^3}}}.
 	\label{Test_Lamb}
 	\end{equation}
 	

 	\section{}
 	\textcolor{black}{
 Here we analyze the integrand of (\ref{mig}) in sub-domain  $ {V_1} = \left\{ {\IB{r} : 1 \le |r| \le \rho } \right\} $,
 where $ \rho $ is an intermediate radius ($ 1 \ll \rho\ll \kappa^{-1} $). 
 	The order of magnitude of the leading order disturbance and test field velocity are:
 	\begin{subequations}\label{OOM_dist}
 		\begin{gather}
 		\IB{v}_{(0)} \sim   H\!a \zeta_{p} O(1/r^{3})  + \beta \, O(1/r^{2}) +  \beta \, O(1/r^{4}) + \gamma \, O(1/r^{3})  + \gamma \, O(1/r^{5}) \mbox{ and\ } \label{OOM_vel}\\
 		\IB{u}^{t} \sim  O(1/r) + O(1/r^{3}). \label{OOM_test}
 		\end{gather}
 	\end{subequations}
 	From (\ref{Vinf0}), we obtain the order of magnitude of the undisturbed velocity
 	\begin{equation}
 	{\IB{U}\!\!_{\infty\, (0)}} \sim  \beta \, O(r) + \gamma \, O(r^{2}) + H\!a \zeta_{p} O(1) + \gamma \, O(1).
 	\label{OOM_Vinf}
 	\end{equation}
 	We substitute (\ref{OOM_dist} \& \ref{OOM_Vinf}) in (\ref{s}) and use it in (\ref{mig}) to find the order of magnitude of integrand $ \bten{s}_{(0)} \colon \bnabla \IB{u}^{t} $ in $ V_{1} $:
 	\small
 	\begin{eqnarray}  	\label{integrand}
 	= \,&& (H\!a \zeta_{p})^{2}\, O\! \left(    \frac{1}{r^{10}} + \delta \frac{1}{r^{7}} + \cdots \right)  
 	+  \beta \, H\!a \zeta_{p} \; O\! \left(\frac{1}{r^{6}}  + \delta \frac{1}{r^{6}} + \cdots  \right)  + \; \gamma \, H\!a \zeta_{p}  \; O\! \left(\frac{1}{r^{5}} + \delta \frac{1}{r^{5}}  + \cdots  \right)  \nonumber\\
 	&& + \, \beta^{2}  \; O\! \left(\frac{1}{r^{5}} + \delta \frac{1}{r^{5}}  + \cdots  \right) + \beta \gamma  \; O\! \left( \frac{1}{r^{4}}  + \delta \frac{1}{r^{6}} + \cdots  \right) + \gamma^{2}  O\! \left( \frac{1}{r^{5}}  + \delta \frac{1}{r^{5}} + \cdots  \right)  .
 	\end{eqnarray}
 	\normalsize
 		The coefficients of different terms in (\ref{integrand}) depict contributions to the integrand arising from various interactions. The radial order describes the range of interaction.
 		Equation (\ref{final}) suggests that the contribution from all the terms in (\ref{integrand}) vanish except the second and fifth term. 
 	}
 
 \textcolor{black}{
We now analyze the error associated with the approximation of the upper limit of integral as $ r = \infty $ in (\ref{final}). Following \cite{ho1974}, the intermediate radius $ \rho $ is written as 
\begin{equation}
\rho \sim O(\kappa^{\eta -1}), \mbox{ where\ }  0< \eta <1.
\end{equation}
Out of the two non-vanishing components of integrand in (\ref{integrand}), the fifth term decays the slowest $ \sim O(1/r^{4}) $.
Thus, the leading order error is generated due to it. Its contribution to the integral ($ \int_{V_1} \bten{s}_{0} : \bnabla{\IB{u}^{t}} dV $) is
\begin{eqnarray}
&=&	\int_{V_{1}:r=1}^{r =\rho \sim O(\kappa^{\eta -1})}  { \beta \gamma O\left(\frac{1}{r^{4}} + \delta \frac{1}{r^{6}} + \cdots \right) } r^{2}  \sin \theta {\rm{d}}\varphi {\rm{d}}\theta {\rm{d}}r    \nonumber \\
&=& 
\beta \gamma \left.O \left( \frac{1}{r} + \delta \frac{1}{r^{3}} + \cdots \right)\right\vert_{r=1}^{r=\rho \sim O(\kappa^{\eta -1}) } =  \beta \gamma \,  \left.O \left( \frac{1}{r} + \delta \frac{1}{r^{3}} + \cdots  \right)\right\vert_{r=1}^{r = \infty}  + \beta \gamma \, O\left(\frac{1}{\kappa^{\eta-1}}\right). \nonumber \\
\end{eqnarray}
Using the definitions in (\ref{alpha}), we conclude that the error incurred above is $ O(\kappa^{2-\eta}) $. For $ 0<\eta<1 $ and $ 1 \ll \rho \ll \kappa^{-1} $, the error is $ o(\kappa) $.
}

 	\section{}
 	
 	\subsection{Validation using the 3-D Giesekus theorem}
 	For a special case of $ \delta=-1 $ (corresponding to $ \Psi_{2}^{*}=-0.5 \, \Psi_{1}^{*} $), Giesekus theorem \citep{g3d,bird1987dynamics} states that the velocity in a second-order fluid is unaltered from the Newtonian velocity. However, the steady pressure field alters as
 	\begin{equation}
 	P = P_{(0)}  +  De \, \delta \, \left( \IB{U}_{(0)} \bcdot \bnabla P_{(0)}  + \frac{1}{4} \left( {\bten{E}^{(1)}_{(0)} \!  \textbf{:} \bten{E}^{(1)}_{(0)}} \right)  \right).
 	\label{G3D_Pressure}
 	\end{equation}
 	We substitute the Newtonian velocity and pressure field and evaluate the modified pressure. This pressure is now substituted in the total hydrodynamic stress tensor 
 	$ \bten{T} = - P \bten{I} + \bten{E}^{(1)} + De \, \bten{S}. $
 	The polymeric stress ($ \bten{S} $) can be represented as an addition of the co-rotational derivative ($ \bten{S}_{C} $) and quadratic component ($ \bten{S}_{Q} $) of the rate of strain tensor \citep{koch2006stress}
 	\small
 	\begin{equation}
 	\bten{S}_{Q} = (1+\delta) \left(  \bten{E}^{(1)} \bcdot \bten{E}^{(1)} \right) \mbox{ and\ } \bten{S}_{C} = \delta \left[ \frac{\p \bten{E}^{(1)}}{\p t}  + \IB{U} \bcdot \bnabla \bten{E}^{(1)} + \frac{1}{2}\left(\bten{R}^{(1)} \bcdot \bten{E}^{(1)}  + \bten{E}^{(1)} \bcdot  \bten{R}^{(1)\,\dagger}\right) \right],
 	\label{Corot_Quadratic}
 	\end{equation}
 	\normalsize
 	where $ \bten{R}^{(1)} $ is the rate of rotation tensor $ (\bnabla \IB{U} - \bnabla \IB{U}^{\dagger}) $. In the limit of $ \delta=-1 $, the quadratic stress vanishes.
 	
 	Accounting for the leading order effects of viscoelasticity on the lateral force (provided $ \delta=-1 $) yields
 	\begin{equation}
 	De \, \IB{F}_{(1)}=\int_{S} De \, \IB{T}_{(1)} \! \bcdot \IB{n} \, {\rm{d}}S = \int_{S} De \, (  -P_{(1)} \bten{I} + \bten{S}_{C \,(0)}  ) \! \bcdot \IB{n} \, {\rm{d}}S.
 	\label{G3D_Force}
 	\end{equation}
 	Here, $ \IB{n} $ is the unit normal vector pointing outside the sphere and steady-state $ \bten{S}_{C \, (0)} $ is evaluated by substituting the Newtonian field ($ \IB{U}_{(0)}, P_{(0)} $) in (\ref{Corot_Quadratic}). Solving the above integral results in the following lateral force
 	\begin{equation}
 	De \, \IB{F}_{(1)} = De \left( - \frac{20 \pi}{3} \beta \gamma \right) \IB{e}_{z},
 	\label{G3D_Force_Sol}
 	\end{equation}
 	which is identical to that obtained from (\ref{final}) for $ \delta=-1 $. In the above calculation, $ De \, \delta(-\frac{24 \pi}{3}  \beta \gamma - 12 \pi  \beta H\!a \zeta_{p})  $ is the contribution from pressure field, and $ De \, \delta (\frac{44 \pi}{3} \beta \gamma + 12 \pi  \beta H\!a \zeta_{p}) $ is the contribution from the co-rotational stress. The analysis here shows that the enhancement in migration comes entirely from the quadratic stress.

 	\subsection{Evaluation and validation of the particle trajectory}
 	\textcolor{black}{
 	The ratio of migration velocity to the translational velocity is
 	\begin{equation}
 		\frac{U_{mig}}{U_{s\, x}}=\frac{ds/dt}{dx/dt}=\frac{ds}{dx}
 		.
 	\end{equation}
 	At the leading order, $ U_{s\,x} $ can be approximated as $ U_{s\,x(0)} $ because the corrections arrive at $ O(De) $ (due to weak non-Newtonian effects) and $ O(\kappa^{2}) $ (due to wall induced viscous resistance). $ U_{mig} $ has both hydrodynamic and Maxwell stress contributions (i.e. $ U_{mig}^{H}+U_{mig}^{M} $).
 	Using first order Euler method and substituting $ U_{mig}^{H} $, $ U_{mig}^{M} $, and $ U_{s\,x(0)} $, we obtain the equation governing the particle trajectory
 	\begin{eqnarray}
 		&&s^{n+1} \approx s^{n} + \Delta x \left( \frac{U_{mig}^{H}+U_{mig}^{M}}{U_{s\, x(0)}}  \right) \nonumber \\
 		&&\approx  s^{n} + \Delta x \left[\frac{ De \left(\frac{10 \pi}{3} \beta \gamma \left ( 1 \! + \! 3 \delta  \right)  \! - \! \frac{3\pi}{2} \beta H\!a \zeta_{p} \left ( 1 \! + \! \delta  \right)\right)     +    4\pi H\!a \left( {\frac{{3\pi }}{{16}}{\kappa ^4}\left(\mathfrak{Z}{\left({4,s} \right) - \mathfrak{Z} \left( {4,1 - s} \right)} \right)} \right)   }{6\pi (1+ O(\kappa)) \left(\alpha+\frac{\gamma}{3} + H\!a (\zeta_{p}-\zeta_{w}) \right)   }\right] . \nonumber\\	\end{eqnarray}
 	Since $ \alpha \sim O(\kappa^{-1}) $, $ \gamma \sim O(\kappa) $, $ H\!a (\zeta_{p}-\zeta_{w}) \sim O(1) $, the trajectory equation can be approximately represented as
 	\begin{eqnarray}\label{traj}
 		s^{n+1} \approx \, s^{n} + &\frac{\Delta x}{6\pi \left(1+O(\kappa) \right)  \alpha}& \left[   De \left(\frac{10 \pi}{3} \beta \gamma \left ( 1 \! + \! 3 \delta  \right)  \! 
 		- \! \frac{3\pi}{2} \beta H\!a \zeta_{p} \left ( 1 \! + \! \delta  \right)\right)   \right. \nonumber \\
 	&& \; \;	\left.
 		    +    4\pi H\!a \left( {\frac{{3\pi }}{{16}}{\kappa ^4}\left(\mathfrak{Z}{\left({4,s} \right) - \mathfrak{Z} \left( {4,1 - s} \right)} \right)} \right)   \right] .
 	\end{eqnarray}
 	We convert this trajectory into an equivalent particle stream width (PSW) for the case of centerline migration (fig.\ref{fig:exp}-b). Since the concentration of suspension is dilute ($ < $0.1$ \% $ in volume fraction), we neglect particle-particle interactions.
 	At the inlet,  particles are homogeneously suspended and cover the entire width of the channel. As the suspension flows, the particles migrate towards the centerline.
 	 	 }

  	 \begin{figure}
  	 	\centering
  	 	\begin{tikzpicture}[baseline]
  	 	\begin{axis}[width=0.4\textwidth,xlabel= $ x^{*} $ (cm), ylabel=PSW, 
  	 	ylabel near ticks,
  	 	xmin=0, xmax=25 , ymin=0, ymax=65, thick, legend style={draw=none,at={(2.6,0.5)},anchor=south east},	legend style={nodes={scale=0.8, transform shape}
  	 		,
  	 		legend columns=1, 
  	 		,	/tikz/column 2/.style={
  	 			column sep=5pt,
  	 		},
  	 	},]
  	 	
  	 	\addplot[olive , mark=*, mark options={scale=0.8}, only marks, error bars/.cd, y dir=both, y explicit] table[x=x, y=PSW] {
  	 		x	PSW
  	 		0.02445504	58.6668507
  	 		1.482864672	50.34449226
  	 		2.712286207	44.68707731
  	 		4.268515998	40.03212166
  	 		5.934793466	35.37799415
  	 		8.817153358	29.06635131
  	 		11.25042979	25.41799566
  	 		13.45916451	23.10128361
  	 		17.10518859	18.79538753
  	 		21.07802092	15.49195076
  	 		24.38612082	13.51684505

  	 	};
  	 	
  	 	\addplot[olive, thick] table[x=x, y=PSW] {Leal.txt};
  	 	\addplot[blue, dash dot, thick] table[x=x, y=PSW-wall] {Leal.txt};
  	 	
  	 	\legend{\citep{ho1976migration}, Present work (without wall correction), Present work (with $ \mathcal{O}(\kappa) $ wall correction)}
  	 	\end{axis}
  	 	\end{tikzpicture}

  	 	\caption{Comparison with \citep{ho1976migration}: particle stream width (PSW) of electrically neutral particles ($ \zeta_{p}=0 $), suspended in a viscoelastic flow. Parameters: $ De=0.12 $, $ \delta=-0.5 $, $ l^{*}= 66 \, \mu $m, $ \kappa =0.017$.}
  	 	\label{fig:Traj_validation}
  	 \end{figure}
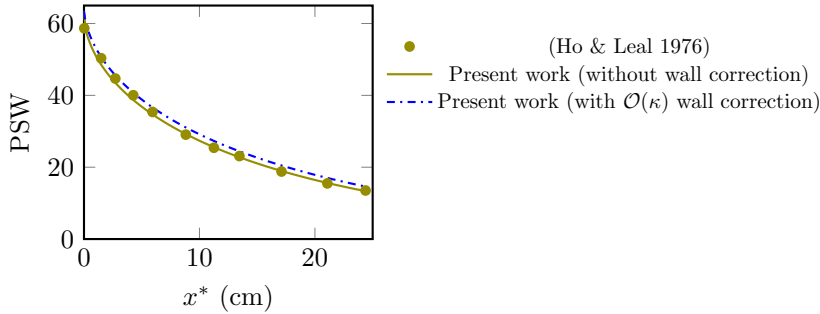

 \textcolor{black}{
 	Experimentally, PSW is measured by tracking the spread of particles (i.e. for centerline focusing, the particle closest to the walls is tracked).
 	Theoretically,  using (\ref{traj}), we find the trajectory of this particle, which is initially suspended near the wall ($ s^{1}=0.95 $). The non-dimensional PSW associated with this trajectory is evaluated using $ |1-2(1-s^{n})| $.  	
 	For instance, $ s^{n} = 0.8 \mbox{ or\ } 0.2 $ corresponds to PSW$ =0.6 $ i.e. the dilute suspension of particles cover 60\% of the channel width.
 	In dimensional form, the particle stream width is $ |l^{*}-2(l^{*}-d^{*\,n})| $. 
 	Fig. \ref{fig:Traj_validation} depicts the stream width of particles suspended in a pressure driven flow of second-order fluid, in the absence of an electric field. The result matches well with that reported by \cite[p.796]{ho1976migration} (the trajectories are converted to PSW using $ |l^{*}-2(l^{*}-d^{*\,n})| $). 
 	The dashed line shows the particle stream width when wall effects are taken into account (at the leading order). Results in \S 4 include this first order wall correction.
}

 	\bibliographystyle{jfm}
 	\bibliography{Akash,randombib}

 \end{document}